\documentclass[12pt]{article}
\usepackage{etex}
\usepackage{epsfig}
\usepackage{epsf}
\usepackage{latexsym}
\usepackage{amsfonts,amssymb,amsmath} 
\usepackage{mathtools}
\usepackage{color}
\usepackage[a4paper,vmargin=3cm,hmargin=2.1cm]{geometry}
\usepackage{tikz}
\epsfverbosetrue
\newcommand{\de}{\hbox{\rm{d}}}

\newcommand{\lb}{\left[}
\newcommand{\rb}{\right]}
\newcommand{\lp}{\left(}
\newcommand{\rp}{\right)}
\newcommand{\la}{\left\{}
\newcommand{\ra}{\right\}}

\newcommand{\dpp}{\vcentcolon}
\newcommand{\bb}{\begin{eqnarray}}
\newcommand{\ee}{\end{eqnarray}}
\newcommand{\eee}{\nonumber\end{eqnarray}}
\newcommand{\qq}{\quad}

\newcommand{\nc}{\newcommand}
 \nc{\alf}{\alpha} \nc{\La}{\Lambda}
  \nc{\ze}{\zeta}
\nc{\tht}{\theta} \nc{\T}{\Theta} \nc{\be}{\beta}  \nc{\eps}{\epsilon} 
\nc{\ga}{\gamma}  \nc{\De}{\Delta} 
 \nc{\G}{\Gamma}  \nc{\vphi}{\varphi}
 \nc{\si}{\sigma}  \nc{\ka}{\kappa}   \nc{\Si}{\Sigma} 
\nc{\om}{\omega}  

\nc{\qqq}{\quad\quad}               

 \nc{\Om}{\Omega}
\nc{\nf}{\infty}   \nc{\dl}{\mathop{\smash{\cal L}}}  \nc{\black}{\rule{3mm}{3mm}}
\nc{\ol}{\overline}        \nc{\und}{\underline} 
\nc{\beq}{\begin{equation}}  \nc{\eeq}{\end{equation}}  \nc{\pt}{\partial}  
   \nc{\dst}{\displaystyle}  \nc{\na}{\nabla} 
\nc{\nnb}{\nonumber}    \nc{\bs}{\backslash}        \nc{\mb}{\mathbb}   
\nc{\sn}{{\rm sn}\,} \nc{\cn}{{\rm cn}\,}     \nc{\dn}{{\rm dn}\,} \nc{\nin}{\noindent}
\nc{\ti}{\tilde}   \nc{\wti}{\widetilde}   \nc{\h}{\hat}  \nc{\wh}{\widehat}
\nc{\tpsi}{\wti{\psi}}   \nc{\tphi}{\wti{\phi}}  \nc{\tH}{\wti{H}} 

\newcounter{muni}
\newenvironment{remunerate}{\begin{list}{{\rm \arabic{muni}.}}
{\usecounter{muni}
\setlength{\leftmargin}{0pt}\setlength{\itemindent}{38pt}}}{\end{list}}

\nc{\brm}{\begin{remunerate}}   \nc{\erm}{\end{remunerate}}

\nc{\stg}{\mathop{\smash{*}}}
\nc{\st}{\mathop{\smash{\delta}}}
\nc{\barr}{\begin{array}}   \nc{\earr}{\end{array}}   \nc{\dg}{\dagger}
\nc{\mtvb}{\mathversion{bold}}   \nc{\mtvn}{\mathversion{normal}}

\begin{document}

\thispagestyle{empty}

\begin{center}
${}$
\vspace{3cm}

{\Large\textbf{Bianchi I meets the Hubble diagram}} \\

\vspace{2cm}

{\large

Thomas Sch\"ucker\footnote{
CPT, Aix-Marseille University, Universit\'e de Toulon, CNRS UMR 7332, 13288 Marseille, France
\\\indent\qq
thomas.schucker@gmail.com },
Andr\'e Tilquin\footnote{CPPM, Aix-Marseille University, CNRS/IN2P3, 13288 Marseille, France\\\indent\qq tilquin@cppm.in2p3.fr },
Galliano Valent\footnote{CPT, Aix-Marseille University, Universit\'e de Toulon, CNRS UMR 7332, 13288 Marseille, France\\\indent\qq
Sorbonne UniversitŽs, UPMC Univ Paris 06, LPTHE CNRS UMR 7589, 75005 Paris, France
 }}

\vspace{3cm}

{\large\textbf{Abstract}}
\end{center}
We improve existing fits of the Bianchi I  metric to the Hubble diagram of supernovae and find an intriguing yet non-significant signal for anisotropy that should be verified or falsified in the near future by the Large Synoptic Survey Telescope. 

 Since the literature contains two different formulas for the apparent luminosity as a function of time of flight in Bianchi I metrics, we present an independent derivation confirming the result by Saunders (1969). The present fit differs from earlier ones by Koivisto \& Mota  and by Campanelli et al. in that we use Saunders' formula, a larger sample of supernovae, Union 2 and JLA, and we use the general Bianchi I metric with three distinct eigenvalues.

\vspace{3.6cm}


\section{Introduction}

We would like to explain our motivation for the present analysis by comparing cosmology with the description of the earth's surface. In good approximation, this surface is maximally symmetric, i.e. a sphere. We observe a breaking of this symmetry of the order of one per mill by the geography, for example by Mount Everest, $8.8\, {\rm km}\cdot 2\pi /(\,40\,000\,{\rm km})\,\approx\, 1.4\cdot10^{-3}.$ Of course we would not try to describe these geographic deviations in terms of a simple geometric model. But there is a second breaking of the maximal symmetry, of the order of 3 per mill, that we call geometric. Indeed this breaking admits a simple geometric description, in terms of an oblate ellipsoid. Our polar radius is about 21.3 km shorter than the equatorial ones, $21.3\, {\rm km}\cdot 2\pi /  (\,40\,000\,{\rm km})\,\approx\, 3.3\cdot10^{-3}.$

The Robertson-Walker metric of the cosmological standard model has maximal spatial symmetry. The Bianchi I metric, that we consider in the following for its calculational simplicity, is obtained from the flat Robertson-Walker metric by giving up the three isotropies. It is true that we observe anisotropies of the order of $10^{-5}$ in the cosmic micro-wave background. But we take these to be geographic deviations and do not try to model them by a simple geometry. Our motivation for using the Bianchi I metric is that it might describe a new breaking of maximal symmetry, of geometric type.

Attempts at deciphering an anisotropy in the Hubble diagram are not new. They come in at least three classes.

The first splits the Hubble diagram in two hemispheres, fits both independently and tries to find a splitting direction in which the two fits differ significantly (Kolatt \& Lahav 2001;
  Schwarz \& Weinhorst 2007;
Antoniou \& Perivolaropoulos 2010;  Kalus et al. 2013;  Yang, Wang \& Chu 2013;
Jimenez, Salzano \& Lazkoz 2014).
 However there is no solution of Einstein's equations compatible with the split.

A second class is similar, but does the fitting with a modified theory of gravity, e.g. based on Finsler geometry, Chang et al. (2014a,b).

The third class is the most conservative. Start with a (pseudo-) Riemannian geometry admitting less symmetries than the Robertson-Walker metric, mostly Bianchi I, compute redshift and apparent luminosity, solve Einstein's equation and then confront this model with the Hubble diagram of supernovae. 

 We are aware of two analyses of this type, by Koivisto \& Mota (2008a) and by Campanelli et al. (2010). Neither finds a preferred direction in the Hubble diagram.

These analyses rest on four main ingredients: two kinematic formulae, the redshift and the apparent luminosity as functions of the three scale factors, and two dynamical formulae, the Einstein equation and its solutions with cosmological
constant and dust. Both analyses (Koivisto \& Mota 2007a;  Campanelli et al. 2010) take the first three ingredients from Koivisto \& Mota (2008b)  and presumably solve the Einstein equation numerically with a Runge-Kutta algorithm. 

All mentioned
 authors seem to be unaware of a work by Saunders (1969) giving all four ingredients. His apparent luminosity and his Einstein equation disagree with the formulae by Koivisto \& Mota (2008b). The disagreement on the apparent luminosity is particularly bothersome: Saunders derives it using a result
 from an earlier paper by himself which in turn relies on a theorem by Ehlers and Sachs. Koivisto \& Mota (2008b) on the other hand derive the apparent luminosity in eight lines.

In the present paper we give {\it ab initio} derivations of the four ingredients. Our luminosity agrees with Saunders' and our Einstein equation agrees with Koivisto \& Mota's. Therefore we redo the fit to the supernovae. We also include recent data and extend the fit to the general Bianchi I metric with three distinct eigenvalues.

 \section{Geodesics}
 
 The Bianchi I metric reads
 \bb \de \tau^2 = \de t^2 - a(t)^2\,\de x^2
 - b(t)^2\,\de y^2- c(t)^2\,\de z^2,\qq a(t),\,b(t),\,c(t)>0.\label{bianchi}\ee
 It is homogeneous but not isotropic. Its non-vanishing Christoffel symbols are:
 \bb \begin{matrix}
 {\Gamma ^t}_{xx}=aa',&
  {\Gamma ^t}_{yy}=bb',&
   {\Gamma ^t}_{zz}=cc',\\[4mm]
  {\Gamma ^x}_{tx}=a'/a,& 
  {\Gamma ^y}_{ty}=b'/b,& 
  {\Gamma ^z}_{tz}=c'/c, 
\end{matrix}
\ee
 where we use $'\dpp=\de/\de t$. Thanks to the isometries under translations, the geodesic equations can be integrated once to read,
 \bb \dot x =A/a^2,\qq \dot y =B/b^2,\qq \dot z=C/c^2,\qq \dot t^2= K+A^2/a^2+B^2/b^2+C^2/c^2,\label{geo}\ee
 where we use $\dot {}\dpp=\de/\de p$ with an affine parameter $p$. $A,\,B,\,C$ and $K$ are integration constants.  We have the time-like solutions $t=p=\tau,\ x=x_0,\ y=y_0,\ z=z_0$ for comoving galaxies, $A=B=C=0,\, K=1$. To describe photons going between them we need light-like geodesics, $K=0$. Let us take the initial conditions at $p=p_{-1}$:
 \bb \begin{matrix}t=t_{-1},&x=x_{-1},&y=y_{-1},&z=z_{-1},\\[6mm]
\dot t=1/W_{-1},&\dot x=A/a^2_{-1},&\dot y=B/b^2_{-1},&\dot z=C/c^2_{-1},
 \end{matrix}\ee
     with the abbreviations $a_{-1}\dpp =a(t_{-1}),\,...$ and 
\bb W(t)\dpp =\lp \,\frac{A^2}{a(t)^2}\, + 
\,\frac{B^2}{b(t)^2}\, +
\,\frac{C^2}{c(t)^2}\, \rp ^{-1/2},\ee
To simplify notations we put $\vec x_{-1}=0$.   
Let us say that the photon arrives today $t=t_0$ at comoving position $\vec x_0$. Then we have:
\bb x_0=\int_{t_{-1}}^{t_0}\,\frac{A}{a^2}\,W\,\de t,\qq y_0=\int_{t_{-1}}^{t_0}\,\frac{B}{b^2}\,W\,\de t,\qq z_0=\int_{t_{-1}}^{t_0}\,\frac{C}{c^2}\,W\,\de t.\label{sol}\ee
In our conventions the speed of light is unity; proper time $\tau$, the coordinate time $t$ and the affine parameter $p$ have units of seconds, the comoving coordinates $x,\,y,\,z$ are dimensionless and the scale factors $a,\,b,\,c$ and the integration constants $A,\,B,\,C$ carry seconds. Then $W$ is dimensionless. Through a rescaling of the affine parameter $p$ we may achieve $A^2+B^2+C^2=1\,{\rm s}^2.$
Through a rescaling of the coordinates $x,\,y,\,z$ we may achieve $a_0=b_0=c_0=1\,{\rm s}.$ 

 We also learnt from equation (\ref{geo}) that in the Bianchi I metric, the position in the sky of  luminous sources changes with time (unless they lie in a principle direction, e.g. $B=C=0$). Since no such drift or `cosmic parallax' (Quercellini, Quartin \& Amendola 2009) has been observed today, any deviation of Bianchi type I from maximal symmetry  must be small (Fontanini, West \& Trodden 2009; Campanelli et al. 2011).

\section{Redshift}

To compute the redshift of the galaxy at $\vec x_{-1}=0$, let it send out a second go-between an atomic period $T$ later, $t=t_{-1}+T$. The atomic period is very small with respect to the time of flight: $T\ll t_0-t_{-1}.$ We want the second go-between to arrive at the same comoving location $\vec x_0$ where our detector sits. This means that we must slightly change the initial direction of the light-like geodesics coded in the integration constants. Let us call them $\tilde A,\,\tilde B,\,\tilde C$. They are close to the former integrations constants:
\bb \tilde A=A(1+\alpha ),\qq
\tilde B=B(1+\beta  ),\qq
\tilde C=C(1+\gamma  ),\qq
\alpha ,\,\beta ,\,\gamma \approx T/(t_0-t_{-1}).\ee
We call the arrival time of the second go-between $t_0+T_D$. In order to compute the Doppler-shifted atomic period $T_D$ at $\vec x_0$ we have to solve the geodesic equation (\ref{geo}) for the second go-between with integration constants $\tilde K=0,\,\tilde A,\,\tilde B,\,\tilde C$ and initial conditions
\bb \dot t(p_{-\tilde1})=\sqrt{\,\frac{\tilde A^2}{a^2_{-\tilde{1}}}\,+\,\frac{\tilde B^2}{b^2_{-\tilde{1}}}\,+\,\frac{\tilde C^2}{c^2_{-\tilde{1}}}\ },\qq \dot x=\,\frac{\tilde A}{a_{-\tilde 1}}\,,\qq \dot y=\,\frac{\tilde B}{b_{-\tilde 1}}\,,\qq  \dot z=\,\frac{\tilde C}{c_{-\tilde 1}}\,,  
\ee
with $a_{-\tilde 1}\dpp=a(t_{-1}+T)$, ... As before the unique solution is:
\bb
x_0=\int_{t_{-1}+T}^{t_0+T_D}\,\frac{\tilde A}{a^2}\, \tilde W\,\de t,\qq
y_0=\int_{t_{-1}+T}^{t_0+T_D}\,\frac{\tilde B}{b^2}\, \tilde W\,\de t,\qq
z_0=\int_{t_{-1}+T}^{t_0+T_D}\,\frac{\tilde C}{c^2}\, \tilde W\,\de t,
\ee
 with 
 \bb \tilde W(t)\dpp =\lp \,\frac{\tilde A^2}{a(t)^2}\, + 
\,\frac{\tilde B^2}{b(t)^2}\, +
\,\frac{\tilde C^2}{c(t)^2}\, \rp ^{-1/2},\ee
To first order in $T/(t_0-t_{-1})$ we get:
\bb x_0&\approx& \int_{t_{-1}}^{t_0}\,\frac{A}{a^2}\,W\,\de t\,+\,\alpha \int_{t_{-1}}^{t_0}\,\frac{A}{a^2}\,\lp\,\frac{B^2}{b^2}\, +
\,\frac{C^2}{c^2}\, \rp W^3\,\de t\,-\,\beta 
\int_{t_{-1}}^{t_0}\,\frac{A}{a^2}\,\frac{B^2}{b^2}\,  W^3\,\de t\,\nonumber\\[2mm]
&&
-\,\gamma  
\int_{t_{-1}}^{t_0}\,\frac{A}{a^2}\,\frac{C^2}{c^2}\,  W^3\,\de t\,-\,\frac{A}{a^2_{-1}}\, W_{-1}T
+\,\frac{A}{a^2_{0}}\, W_{0}T_D\,.
\ee
Together with the first of equations (\ref{sol}) we have:
\bb 0&\approx& \alpha \int_{t_{-1}}^{t_0}\,\frac{W^3}{a^2}\,\lp\,\frac{B^2}{b^2}\, +
\,\frac{C^2}{c^2}\, \rp \,\de t\,-\,\beta 
\int_{t_{-1}}^{t_0}\,\frac{W^3}{a^2}\,\frac{B^2}{b^2}\,\de t\,\nonumber\\[2mm]
&&
-\,\gamma  
\int_{t_{-1}}^{t_0}\,\frac{W^3}{a^2}\,\frac{C^2}{c^2}\,\de t\,-\,\frac{W_{-1}}{a^2_{-1}}\, T
+\,\frac{W_{0}}{a^2_{0}}\, T_D\,\nonumber\\[2mm]
&&= B^2\,Z+C^2\,Y -\,\frac{W_{-1}}{a^2_{-1}}\, T
+\,\frac{W_{0}}{a^2_{0}}\, T_D\,,
\label{x}
\ee 
with
\bb X\dpp=(\beta  -\gamma )\int_{t_{-1}}^{t_0}\,\frac{W^3}{b^2c^2}\,\de t,\qq
Y\dpp=(\alpha -\gamma )\int_{t_{-1}}^{t_0}\,\frac{W^3}{c^2a^2}\,\de t,\qq
Z\dpp=(\alpha -\beta )\int_{t_{-1}}^{t_0}\,\frac{W^3}{a^2b^2}\,\de t. \label{compare}\ee
Similarly for $y_0$,
\bb 0&\approx& -A^2\,Z+C^2\,X-\,\frac{W_{-1}}{b^2_{-1}}\, T
+\,\frac{W_{0}}{b^2_{0}}\, T_D\,,\label{y}
\ee
and for $z_0$,
\bb 0&\approx& -A^2\,Y-B^2\,X-\,\frac{W_{-1}}{c^2_{-1}}\, T
+\,\frac{W_{0}}{c^2_{0}}\, T_D\,.\label{z}
\ee
Solving the three linear equations (\ref{x},\,\ref{y},\,\ref{z}) in $X,\,Y,\,Z$ we remain with
\bb\,\frac{T}{W_{-1}}\, =\,\frac{T_D}{W_0}\, .\ee
Therefore the redshift is:
\bb z\dpp=\,\frac{T_D-T}{T}\, \approx\,\frac{W_0}{W_{-1}}\, -1\,=\,\frac{\sqrt{\,\frac{ A^2}{a^2_{-{1}}}\,+\,\frac{ B^2}{b^2_{-{1}}}\,+\,\frac{ C^2}{c^2_{-{1}}}\ }}{\sqrt{\,\frac{ A^2}{a^2_{{0}}}\,+\,\frac{ B^2}{b^2_{{0}}}\,+\,\frac{ C^2}{c^2_{{0}}}\ }}\,-1\,. 
 \ee
 This formula agrees with the redshift derived in 
 Saunders (1969), Koivisto \& Mota (2008b) and Fontanini et al. (2009).

\section{Apparent luminosity}

To compute the apparent luminosity of a supernova at $\vec x_{-1}$, let it send out a rectangular beam of photons. One corner of this sequence of infinitesimal rectangles is given by the initial direction $\dot x(p_{-1}) =A/a_{-1}^2,\ \dot y(p_{-1}) =B/b_{-1}^2,\ \dot z(p_{-1})=C/c_{-1}^2$. As before this photon leaves the supernova at $t_{-1}$. Formally this initial direction defines a space-like vector $(0,\,A/a_{-1}^2,\, B/b_{-1}^2,\,C/c_{-1}^2)=\dpp (0,\,\vec v_{-1})$ in the tangent space at $(t_{-1},\,\vec x_{-1})$.

 The two adjacent corners are defined by light-like geodesics with initial directions $\vec v_{-1}+\vec\epsilon_{-1}$ and $\vec v_{-1}+\vec\delta _{-1}$ where 
 \bb
 \vec\epsilon_{-1}&\dpp =&\epsilon \lp \,\frac{B}{a_{-1}b_{-1}}\, ,\,\frac{-A}{a_{-1}b_{-1}}\, ,\,\qq
0\rp \\[3mm]
 \vec\delta _{-1}&\dpp =&\delta \lp \,\frac{AC}{a_{-1}^2c_{-1}}\, ,\,\frac{BC}{b_{-1}^2c_{-1}}\, ,\,
-\lp\,\frac{A^2}{a_{-1}^2c_{-1}}\, +\,\frac{B^2}{b_{-1}^2c_{-1}}\,\rp\rp.\ee
We assume that $A$ and $B$ are not both zero.
The three vectors $\vec v_{-1},\,\vec\epsilon_{-1}$ and $\vec\delta_{-1} $ are mutually orthogonal with respect to the metric (\ref{bianchi}). We take $\epsilon$ and $\delta $ dimensionless. Then $\vec v_{-1},\,\vec\epsilon_{-1}$ and $\vec\delta_{-1} $ have units s$^{-1}$. Their norms $|\vec v_{-1}|,\,|\vec\epsilon_{-1}|,\,|\vec\delta_{-1}| $ with respect to the metric are however dimensionless. The solid angle in the rest frame of the supernova at $\vec x_{-1}$ cut out by the rectangular beam is given by
\bb \Omega _{-1}=\,\frac{|\vec\epsilon_{-1}|\cdot|\vec\delta_{-1}|}{4\pi\,|\vec v_{-1}|^2 }\,=\,\,\frac{\epsilon\delta}{4\pi }\,  W_{-1}\lp\frac{A^2}{a_{-1}^2}\, +\,\frac{B^2}{b_{-1}^2}\rp\, .\ee

On its way to the detector at $\vec x_0$ the rectangular beam gets deformed into a sequence of infinitesimal parallelograms.
The final parallelogram is defined by two infinitesimal vectors $\vec\epsilon_0\dpp=\vec x_0^\epsilon-\vec x_0$ and $\vec\delta _0\dpp=\vec x_0^\delta -\vec x_0$, where $\vec x^\epsilon$ is the space part of the light-like geodesic with initial direction $\vec v_{-1}+\vec \epsilon_{-1}$ and likewise for $\delta $. This final parallelogram is the effective surface of the detector and our task is to compute its area. We have
 \bb
x_0^\epsilon&=&\int_{t_{-1}}^{t_0}\,\frac{A+\epsilon B\,\frac{a_{-1}}{b_{-1}}\, }{a^2}\, 
\lp\,\frac{\lp\,{A+\epsilon B\,\frac{a_{-1}}{b_{-1}}}\,\rp^2}{a^2}\, + 
\,\frac{\lp\,{B-\epsilon A\,\frac{b_{-1}}{a_{-1}}}\,\rp^2}{b^2}\, + \,\frac{C^2}{c^2}\, \rp^{-1/2}\,\de t\\[3mm]
&\approx& x_0\,+\,\epsilon B \int_{t_{-1}}^{t_0}
\,\frac{W}{a^2}\, \la \,\frac{a_{-1}}{b_{-1}}\,-\,A^2W^2\,\lb\,\frac{a_{-1}}{b_{-1}}\,\frac{1}{a^2}\, -\,\frac{b_{-1}}{a_{-1}}\,\frac{1}{b^2}\, 
\rb\ra \,\de t  +O(\epsilon^2)\,,\\[5mm]
y_0^\epsilon&=&\int_{t_{-1}}^{t_0}\,\frac{B-\epsilon A\,\frac{b_{-1}}{a_{-1}}\, }{b^2}\, 
\lp\,\frac{\lp\,{A+\epsilon B\,\frac{a_{-1}}{b_{-1}}}\,\rp^2}{a^2}\, + 
\,\frac{\lp\,{B-\epsilon A\,\frac{b_{-1}}{a_{-1}}}\,\rp^2}{b^2}\, + \,\frac{C^2}{c^2}\, \rp^{-1/2}\,\de t\\[3mm]
&\approx& y_0\,+
\,\epsilon A \int_{t_{-1}}^{t_0}
\,\frac{W}{b^2}\la \,\frac{-b_{-1}}{a_{-1}}\,-\,B^2W^2\lb\,\frac{a_{-1}}{b_{-1}}\,\frac{1}{a^2}\, -\,\frac{b_{-1}}{a_{-1}}\,\frac{1}{b^2}\, 
\rb\ra\,\de t
 \qq +O(\epsilon^2)\,\\[5mm]
z_0^\epsilon&=&\int_{t_{-1}}^{t_0}\,\frac{C}{c^2}\, 
\lp\,\frac{\lp\,{A+\epsilon B\,\frac{a_{-1}}{b_{-1}}}\,\rp^2}{a^2}\, + 
\,\frac{\lp\,{B-\epsilon A\,\frac{b_{-1}}{a_{-1}}}\,\rp^2}{b^2}\, + \,\frac{C^2}{c^2}\, \rp^{-1/2}\,\de t\\[3mm]
&\approx& z_0\,-\,\epsilon 
ABC \int_{t_{-1}}^{t_0}
\,\frac{W^3}{c^2} \lb\,\frac{a_{-1}}{b_{-1}}\,\frac{1}{a^2}\, -\,\frac{b_{-1}}{a_{-1}}\,\frac{1}{b^2}\, 
\rb\,\de t \qq +O(\epsilon^2)\,.\ee
Similarly, one computes the components in $\delta $ and obtains to first order in $\epsilon $ and $\delta $:
\bb 
\vec\epsilon_0&\approx&\epsilon\begin{pmatrix}
 B \int_{t_{-1}}^{t_0}
\,\frac{W}{a^2} \la \,\frac{a_{-1}}{b_{-1}}\,-\,A^2W^2\lb\,\frac{a_{-1}}{b_{-1}}\,\frac{1}{a^2}\, -\,\frac{b_{-1}}{a_{-1}}\,\frac{1}{b^2}\, 
\rb\ra\,\de t\\[3mm]
 A \int_{t_{-1}}^{t_0}
\,\frac{W}{b^2}\la \,\frac{-b_{-1}}{a_{-1}}\,-\,B^2W^2\lb\,\frac{a_{-1}}{b_{-1}}\,\frac{1}{a^2}\, -\,\frac{b_{-1}}{a_{-1}}\,\frac{1}{b^2}\, 
\rb\ra\,\de t\\[3mm]
 -ABC \int_{t_{-1}}^{t_0}
\,\frac{W^3}{c^2} \lb\,\frac{a_{-1}}{b_{-1}}\,\frac{1}{a^2}\, -\,\frac{b_{-1}}{a_{-1}}\,\frac{1}{b^2}\, 
\rb\,\de t
\end{pmatrix},\\[4mm]
\vec\delta _0&\approx&\delta \begin{pmatrix}
 \,\frac{AC}{c_{-1}}\, \int_{t_{-1}}^{t_0}\,\frac{W}{a^2} \la\,1\,-\,W^2\lb\,\frac{A^2}{a^2}\,+ 
 \,\frac{B^2}{b^2}\,-\,\frac{c_{-1}^2}{c^2} 
 \lp\,\frac{A^2}{a_{-1}^2}\,+ \,\frac{B^2}{b_{-1}^2}\,\rp\rb\ra\,\de t\\[3mm]
 \,\frac{BC}{c_{-1}}\, \int_{t_{-1}}^{t_0}\,\frac{W}{b^2} \la\,1\,-\,W^2\lb\,\frac{A^2}{a^2}\,+ 
 \,\frac{B^2}{b^2}\,-\,\frac{c_{-1}^2}{c^2} 
 \lp\,\frac{A^2}{a_{-1}^2}\,+ \,\frac{B^2}{b_{-1}^2}\,\rp\rb\ra\,\de t\\[3mm]
 - \int_{t_{-1}}^{t_0}\,\frac{W}{c^2} \la\,c_{-1}\lp\frac{A^2}{a_{-1}^2}\,+ \,\frac{B^2}{b_{-1}^2}\rp\,+\,\frac{C^2}{c_{-1}}\, W^2\lb\,\frac{A^2}{a^2}\,+ 
 \,\frac{B^2}{b^2}\,-\,\frac{c_{-1}^2}{c^2} 
 \lp\frac{A^2}{a_{-1}^2}\,+ \,\frac{B^2}{b_{-1}^2}\rp\rb\ra\,\de t
 \end{pmatrix}.\nonumber
\ee
Note that in contrast to our definitions at $t_{-1}$, the vectors
$\vec\epsilon_{0}$ and $\vec\delta_{0} $ are dimensionless and their norms $|\vec\epsilon_{0}|,\,|\vec\delta_{0}| $ with respect to the metric carry seconds. 

To first order in $\epsilon $ and $\delta $ these two vectors are orthogonal to $\vec v_0$ with respect to the metric (\ref{bianchi}):
\bb \vec\epsilon_0\cdot \vec v_0\approx \vec\delta _0\cdot \vec v_0\approx 0.\ee
Therefore the effective area is 
\bb S_0\approx|\vec\epsilon_0\wedge\vec\delta _0|,\ee
where the vector product is computed with the metric (\ref{bianchi}). Finally the apparent luminosity is:
\bb \ell\,=\,L\,\frac{\Omega _{-1}}{S_0 }\lp\,\frac{W_{-1}}{W_0}\,\rp^2, 
\ee
where $L$ denotes the absolute luminosity of the standard candle, that we suppose to be radiating isotropically in its rest frame. Note that in the apparent luminosity $\epsilon $ and $\delta $ cancel.

Our formula disagrees with the one derived in eight lines by Koivisto \& Mota (2008b):
\bb \ell\,=\,\frac{L}{4\pi }\lb\int_{t_{-1}}^{t_0}\,
\lp A^2a^2+B^2b^2+C^2c^2\rp^{-1/2}\,\de t\rb^{-2}
\lp\,\frac{W_{-1}}{W_0}\,\rp^2
\ee
(in our notations and using their normalisations:  $a_0=b_0=c_0=1=(A^2+B^2+C^2)^{1/2})$. If the photon moves in a principle direction, say $B=C=0$, then Koivisto and Mota's apparent luminosity does not distinguish the Minkowskian metric, $a=b=c=1$ s, from the one expanding in the $x$ and $y$ directions, say $ a=b=\exp(Ht)\,{\rm s},\ c=1\,{\rm s}$. This is not compatible with the drift of luminous sources.

Our formula agrees with the one derived by Saunders (1969):
\bb \ell=\,\frac{L}{4\pi }\, \frac{  W_{-1}^5}{a_{-1}b_{-1}c_{-1}\,\Delta \,W_0^4}\, ,\ee
with:
\bb I_x\dpp=-\int_{t_{-1}}^{t_0}\,\frac{W^3}{b^2c^2}\,  \de t,
\qq
I_y\dpp=-\int_{t_{-1}}^{t_0} \,\frac{W^3}{c^2a^2}\,\de t,
\qq
I_x\dpp=-\int_{t_{-1}}^{t_0} \,\frac{W^3}{a^2b^2}\,\de t,
\ee 
(compare with equations (\ref{compare})) and 
$\Delta \dpp = A^2I_yI_z +B^2I_zI_x +C^2I_xI_y$.

\section{Einstein's equations}

To keep things simple, we solve Einstein's equations with a positive cosmological constant $\Lambda$ and dust whose mass density is denoted as usual by $\rho (t)$:
\beq {\rm Ric}_{\mu \nu}-{\textstyle\frac{1}{2}} R\,g_{\mu \nu}=\Lambda \,g_{\mu \nu}+8\pi\,G\,\rho\,{\delta ^0}_\mu\,{\delta ^0}_\nu.  \eeq
Defining the Hubble parameters 
\beq\label{defH}
H_x\dpp=\frac{a'}{a},\qqq H_y\dpp=\frac{b'}{b},\qqq H_z\dpp=\frac{c'}{c}, \qqq {\rm with}\qqq
'\dpp=\,\frac{\de}{\de t}\, ,
\eeq
we have the following differential system:
\beq\label{Einstein}\barr{lrcl}
(a)\qqq & H'_x+H'_y+H_x^2+H_y^2+H_x\,H_y & = & \Lambda,  \\[4mm] 
(b)\qqq & H'_y+H'_z+H_y^2+H_z^2+H_y\,H_z & = & \Lambda,  \\[4mm] 
(c)\qqq & H'_z+H'_x+H_z^2+H_x^2+H_z\,H_x & = & \Lambda,  \\[4mm] 
(d)\qqq & H_x\,H_y+H_y\,H_z+H_z\,H_x     & = & \dst \Lambda +8\pi\,G\,\rho. \earr 
\eeq
to which we may add the covariant energy-momentum conservation which integrates to
\beq
\rho(t)=\frac{r_*}{V(t)} \qqq {\rm with} \qqq V(t)\dpp=
\,a(t)b(t)c(t).  \eeq 
Here $r_*=\dpp \rho _0a_0b_0c_0$ is an integration constant.

\subsection{ The  isotropic case}  
Let us observe that if we take $a=b=c$ we do recover Friedman's equations in the form
\beq\label{Fried}
2H'+3H^2=\La,  \qq\qq  3H^2=\La+8\pi G\rho \qq\qq,\qq H:=\frac{a'}{a}.
\eeq 
For positive $\Lambda $ we integrate the first equation in $H$ and obtain a bifurcation:
\beq \barr{ll}\dst 
 \quad \Lambda /(3H^2)< 1: \qq & \dst \qq H=\sqrt{\frac{\La}{3}}\,\coth(u/2), 
\\[4mm]\dst
  \quad \Lambda /(3H^2)> 1: \qq & \dst \qq H=\sqrt{\frac{\La}{3}}\,\tanh(u/2), \earr 
\qq\qq   
\eeq
where $u\dpp=\sqrt{3\La}\,(t-t_*)$ and $t_*$ is an integration constant. Writing the scale factor $a=a_*\,V^{1/3}$ we have 
\beq\barr{ll}
\dst   \quad \Lambda /(3H^2)< 1 : & \qq\qq\dst 
\frac V{V_0}=+\frac{4\pi G \rho_0}{\La}(\cosh u-1),\\[4mm]
\dst  \quad \Lambda /(3H^2)> 1: \quad & \qq\qq\dst 
\frac V{V_0}=-\frac{4\pi G \rho_0}{\La}(\cosh u+1).\earr 
\eeq
The bifurcation point yields the singular solution $H=\sqrt{\Lambda /3}$, for which the matter density vanishes, $\rho =0$, by the second Friedman equation (\ref{Fried}). The second branch, $\Lambda /(3H^2)>1$, has negative matter density  and no conventional physical interpretation.

For negative cosmological constant, $\La<0$, there is no bifurcation and we get
\beq
H=-\sqrt{\frac{|\La|}{3}}\,\tan(u/2),\qq\qq\qq\qq \frac V{V_0}=\frac{4\pi G \rho_0}{|\La|}\,(1+\cos u).
\eeq

\subsection{Integration of the Einstein equations}

Let us begin with the first three equations: taking the difference between equations $(a)$ and $(b)$ 
we have
\beq\label{Hy}
\frac{(H_y-H_x)'}{(H_y-H_x)}=-H_x-H_y-H_z=-\frac{V'}{V}\, \qq \Longrightarrow\qq H_y=H_x+\frac{L}{V}\,.
\eeq
The same treatment applied to the equations $(a)$ and $(c)$ gives
\beq\label{Hz}
\frac{(H_z-H_x)'}{(H_z-H_x)}=-H_x-H_y-H_z=-\frac{V'}{V}\, \qq \Longrightarrow\qq H_z=H_x+\frac{M}{V}\,.
\eeq 
In this step we got two new integrations constants $L$ and $M$. Combining equations $(b)$ and $(c)$ gives no new relation. 
 
Inserting the previous relations into equations $(a)$, $(b)$ and $(c)$ we get three first order differential 
equations mixing $H_x$ and $V$:
\beq\barr{lrcl}
(a') \qqq & \dst 2H'_x+3H_x^2+\frac{3L}{V}H_x+\frac{L^2-L\,V'}{V^2} & = & \Lambda,  \\[4mm]\dst 
(b') \qqq & \dst 2H'_x+3H_x^2+\frac{3(L+M)}{V}H_x+\frac{L^2+M^2+LM-(L+M)\,V'}{V^2} & = & \Lambda,  \\[4mm]\dst 
(c') \qqq & \dst 2H'_x+3H_x^2+\frac{3M}{V}H_x+\frac{M^2-M\,V'}{V^2} & = & \Lambda . \earr
\eeq
Subtracting $(a')$ from $(b')$ or $(a')$ from $(c')$ yields a single relation for $H_x$, 
\beq\label{fctsH1}
H_x=\frac{V'}{3V}-\frac{L+M}{3V}
\eeq
which, upon use of (\ref{Hy}) and (\ref{Hz}), implies
\beq\label{fctsH2}
H_y=\frac{V'}{3V}+\frac{2L-M}{3V},\qqq 
H_z=\frac{V'}{3V}+\frac{-L+2M}{3V}.
\eeq
We are left with $(a')$  involving only the function $V(t)$. It reads
\beq
2V\,V''-(V')^2-3\Lambda \,V^2 +\si^2=0, \qqq\quad \si^2\dpp=L^2-LM+M^2\geq 0,
\eeq
and is readily integrated once:
\beq
(V')^2=\si^2+2E\,V+3\,\Lambda \,V^2
\eeq
where $E$ is a new integration constant.

It remains just to check the last equation (\ref{Einstein}) $(d)$: the computation gives the very simple relation 
\beq E=12\pi\,G\,r_*.\eeq
The solution for the scale factors $a(t),\,b(t),\,c(t)$ is then obtained in the following steps: 
\brm
\item First, compute the volume $V(t)$ by solving 
\beq\label{mastereq}
(V')^2=\si^2+24\pi\,G\,r_*\,V+3\La\,V^2,\qqq \si^2\dpp=L^2-LM+M^2\geq 0, \qqq \La>0. 
\eeq 
\item Deduce the Hubble parameters $H_x,\,H_y,\,H_z\,$ from equations  (\ref{fctsH1}) and (\ref{fctsH2}). 
\item Finally integrate the Hubble parameters (\ref{defH}) and obtain the scale factors $a,\,b,\,c$.\erm

Let us begin by integrating equation (\ref{mastereq}),
\beq\label{odeV}
\frac{\de V}{\sqrt{\si^2/3\La+2r\,V+V^2}}=\pm\,\sqrt{3\La}\,\de t\,,\qq\qq r:=\frac{4\pi Gr_*}{\La}
\eeq
taking, without loss of generality, the positive sign. It is convenient to define
\beq\label{defxi}
\xi\dpp=\sqrt{\frac{\La}{3}}\frac{\si}{4\pi Gr_*}>0,
\eeq
in order to write
\beq \si^2/3\La+2r\,V+V^2=(V+r)^2-r^2(1-\xi^2). \eeq
The integration of (\ref{odeV}) is then obvious for $\xi=1$ and in the other cases the changes of variables
\bb\xi<1:\qq V+r=r\sqrt{1-\xi^2}\,\cosh u,\qq\qq\qq \xi>1:\qq V+r=r\sqrt{\xi^2-1}\,\sinh u,\ee
give easily the function $V,$ which exhibits a bifurcation parametrized by $\xi$:
\beq\label{fctV}\barr{lcl}
\xi <1 \quad  & :\qqq & \dst 
V=\frac{4\pi G r_*}{\La}\,\Big(\sqrt{1-\xi^2}\,\cosh u-1\Big),
\\[5mm]
\xi=1 \quad  & :\qqq & \dst V=\,\frac{4\pi G r_*}{\La}(e^u-1),
\\[5mm]
\dst \xi>1 \quad  & :\qqq & \dst 
V=\,\frac{4\pi G r_*}{\La}\,\Big(\sqrt{\xi^2-1}\,\sinh u-1\Big),
\earr\qq \eeq   
with $u=\sqrt{3\La}\,(t-t_*)$ as before. 
The very simple structure of these functions may be understood by differentiating equation 
(\ref{mastereq}) which gives the linear equation
\beq 
V''-3\La\, V=12\pi G r_*.\eeq 

For the third step, we show the computation only for $a(t)$ with $\xi <1$. We start from equation
(\ref{fctsH1}),
\beq
H_x=\frac{a'}{a}=\frac{V'}{3V}-\frac{(L+M)}{3V}\,,
\eeq
and integrate:
\beq
a(t)=a_*\,V^{1/3}\,\exp\left(-\frac{(L+M)}{3}\int\frac{\de t}{V}\right).
\eeq
We need to fix the sign of $V$, which is positive. Therefore 
$u>{\rm Artanh}\,\xi$. We have
\beq
\int\frac{\de t}{V}=\frac{\xi }{\si\sqrt{1-\xi ^2}}\int\frac{\de u}{\cosh u-1/\sqrt{1-\xi ^2}}=
\,\frac{2}{\sigma }\, \xi \sqrt{1-\xi ^2}\int \frac{e^u\,\de u}{\lp \sqrt{1-\xi ^2}\,e^u-1\rp^2-\xi ^2}\,,\label{sigma}
\eeq
and the change of variables
\bb v=\,\frac{1}{\xi }\,\lp \sqrt{1-\xi ^2}\,e^u\,-1\rp >1,\ee
yields
\bb
\int\frac{\de t}{V}=
\,\frac{2}{\sigma }\, \int\,\frac{\de v}{v^2-1}\, =
\frac 1{\si}\,\ln\frac{e^u-w}{e^u-1/w},\qq {\rm with}\qq w\dpp=\exp{\rm Artanh}\,\xi =
\sqrt{\frac{1+\xi}{|1-\xi|}}.\ee
 The computations are similar for the
other scale factors. With the definitions
\beq
s_1\dpp=-\frac{L+M}{3\si}, \qqq\qqq s_2\dpp=\frac{2L-M}{3\si}\,,
\eeq
we obtain
\beq
a(t)=a_*\,V(t)^{1/3}\,R(t)^{s_1},\qqq 
b(t)=b_*\,V(t)^{1/3}\,R(t)^{s_2},\qqq 
c(t)=\frac{V(t)}{a(t)\,b(t)},\label{scalepos}
\eeq
where $a_* $ and $b_* $ are integration constants and $R$ is the strictly positive function  of time defined by
\beq 
\left\{\barr{lll}\dst
\xi<1 \  & : \dst\quad  R(t):=\frac{e^u-w}{e^u-1/w}, \quad &  u>{\rm Artanh}\,\xi,
\\[4mm] \xi=1 \  & : \quad R(t):=1-e^{-u}, & u>0,   \\[1mm] 
\dst \xi>1 \ & : \dst \quad R(t):=\frac{e^u-w}{e^u+1/w}, &  u>{\rm Artanh}\,(1/\xi).\qqq  \dst \earr\right.
\eeq

A few remarks are in order:
\begin{itemize}
\item As time grows, the anisotropy of the Universe fades away since $R\to 1$. This is a well-known property 
of Bianchi I cosmologies. In particular, equations (\ref{Hy}) and  (\ref{Hz}) show that this fading is driven by increasing volume $V$.
\item
 If we choose $t_* =-({\rm Artanh}\,\xi)/\sqrt{3\Lambda }$ in the first case, $\xi <1$, then the big bang occurs at $t=0$.
\item
The discussed solution is usually attributed to Saunders (1969). His field equations 
are
\beq
\barr{lcl} H_x'+\frac 13\,H_x(H_x+H_y+H_z) & = & \La+4\pi G\rho,\\[4mm]
H_y'+\frac 13\,H_y(H_x+H_y+H_z) & = & \La+4\pi G\rho,\\[4mm]
H_z'+\frac 13\,H_z(H_x+H_y+H_z) & = & \La+4\pi G\rho,\\[4mm]
H_x\,H_y+H_y\,H_z+H_z\,H_x & = & \La+8\pi G\rho.\earr\eeq
In the isotropic limit where $H_x=H_y=H_z=\dpp H$ we obtain
\beq
H'+H^2=\La+4\pi G\rho \qq\qq 3H^2=\La+8\pi G\rho.
\eeq
Eliminating $\rho$ in the first equation by using the second equation they read
\beq
2H'-H^2=\La\qq\qq 3H^2=\La+8\pi G\rho\eeq
and are at variance with Friedman's equations (\ref{Fried}).
\item
Note that in the isotropic limit, $\sigma $ goes to zero and some intermediate results, e.g. equation (\ref{sigma}), in our derivation of the solutions to Einstein's equations are singular. One way to avoid these singularities
is to take the limit in two steps with the intermediate step being the ellipsoid of revolution. In any case, our final solutions, the scale factors (\ref{scalepos}), have a well defined limit.

\item 
For completeness we also indicate the case of negative cosmological constant.
Defining this time 
\bb r:=\frac{4\pi G r_*}{|\La|} \qq\qq \qq \xi\dpp=\sqrt{\frac{|\La|}{3}}\frac{\si}{4\pi Gr_*},\ee
we can write the differential equation (\ref{mastereq}) for $V$ in the form
\beq
\frac{\de V}{\sqrt{-(V+r)^2+r^2(1-\xi^2)}}=\sqrt{3|\La|}\,\de t\eeq
showing that there is no bifurcation since now $\xi$ must be smaller than one for $V$ to remain real. The 
change of variables
\beq
V+r=r\sqrt{1-\xi^2}\,\sin u\eeq
gives easily
\beq
\frac V{V_0}=-\frac{4\pi G\rho_0}{|\La|}\Big(1-\sqrt{1-\xi^2}\,\sin u\Big)<0 
\qq\qq\qq u=\sqrt{3|\La|}\,(t-t_*),
\eeq
and if one defines
\bb s_1=2\,\frac{L+M}{3\si},\qq\qq\qq s_2=-2\,\frac{2L-M}{3\si},\ee
the scale factors are
\bb
a(t)&=&a_*\,V(t)^{1/3}\,\exp[s_1\,\arctan R(t)],\qq\\[2mm] 
 b(t)&=&b_*\,V(t)^{1/3}\,\exp[s_2\,\arctan R(t)], \qqq
c(t)\ =\ \frac{V(t)}{a(t)\,b(t)},
\ee
with
\beq
R(t)=\frac{\arctan(u/2)-\sqrt{1-\xi^2}}{\xi}.
\eeq
\end{itemize}

 \section{Small eccentricities}

No drift in the positions of quasars or galaxies has been observed today. We therefore assume that the three scale factors $a,\,b$ and $c$ differ only by small amounts:
\bb b(t)=\dpp a(t)\lb1+\beta (t)\rb,\qq 
c(t)=\dpp a(t)\lb1+\eta (t)\rb, \qq \beta,\eta\ll 1.\ee
 As explained after equation (\ref{sol}), we may set $\beta(t_0)=\eta(t_0)=0$ without loss of generality.  
 
 \subsection{Kinematics}
 
 In the following we keep only leading terms in $\epsilon,\,\delta$ and $\beta ,\,\eta$, and continue using $\approx$ to indicate this approximation. 

Let us introduce the abbreviations $N^2\dpp=A^2+B^2+C^2=1\,{\rm s}^2$,   $\chi \dpp =\int_{t_{-1}}^{t_0}\de t/a$, $\bar\beta \dpp =\chi ^{-1}\int_{t_{-1}}^{t_0}\beta\,\de t/a$ and $\bar\eta \dpp =\chi ^{-1}\int_{t_{-1}}^{t_0}\eta\,\de t/a$. Note that $\chi $ is the dimensionless comoving geodesic ``distance'' between the supernova emitting the photon
 and the Earth. In these notations the kinematics reads:
\bb W&\approx &\frac{a}{N}\,\lb 1+\,\frac{B^2}{N^2}\,  \beta+\,\frac{C^2}{N^2}\,  \eta\rb,\\[3mm]
z+1&= &\frac{W_0}{W_{-1}}\,\approx \,\frac{a_0}{a_{-1}}\lb1-\,\frac{B^2}{N^2}\,  \beta_{-1}-\,\frac{C^2}{N^2}\,\eta_{-1}\rb,\label{red}
\ee
\bb
\epsilon_{0x}&\approx&\epsilon\,\frac{B}{N}\,\chi \lb
1-\lp 1-2\,\frac{A^2}{N^2}\,\rp\beta _{-1}+
\lp-2+3  \,\frac{B^2}{N^2}\,+2\,\frac{C^2}{N^2}\,\rp
\bar\beta 
+\,\frac{C^2}{N^2}\,\bar\eta\rb,
\\[3mm]
\epsilon_{0y}&\approx&-\,\epsilon\,\frac{A}{N}\,\chi \lb
1+\lp 1-2\,\frac{\beta ^2}{N^2}\,\rp\beta _{-1}+
\lp-2+3  \,\frac{B^2}{N^2}\,\qqq\ \rp
\bar\beta 
+\,\frac{C^2}{N^2}\,\bar\eta\rb,
\\[3mm]
\epsilon_{0z}&\approx&-\,\epsilon\,\frac{ABC}{N^3}\,\chi \lb
-2\,\beta _{-1}+2\,
\bar\beta\, \rb,
\\[3mm]
\delta _{0x}&\approx&\delta \,\frac{AC}{Na_{-1}}\,\chi \lb
1-2\,\frac{B^2}{N^2}\,\beta _{-1}+
3  \,\frac{B^2}{N^2}\,
\bar\beta +
\lp 1-2\,\frac{C^2}{N^2}\,\rp\eta _{-1}+
\lp-2+3  \,\frac{C^2}{N^2}\,\rp
\bar\eta \rb,
\\[3mm]
\delta _{0y}&\approx&\delta \,\frac{BC}{Na_{-1}}\,\chi \lb
1-2\,\frac{B^2}{N^2}\,\beta _{-1}+
\lp-2+3  \,\frac{B^2}{N^2}\,\rp
\bar\beta \right. \nonumber\\
&&\qqq\qqq\qqq\qqq\qqq\qqq\qqq\ \ \left.+
\lp 1-2\,\frac{C^2}{N^2}\,\rp\eta _{-1}+
\lp-2+3  \,\frac{C^2}{N^2}\,\rp
\bar\eta \rb,
\\[3mm]
\delta _{0z}&\approx&-\,\delta \,\frac{A^2+B^2}{Na_{-1}}\,\chi \lb
1-2\,\frac{B^2}{N^2}\,\beta _{-1}+
\,\frac{B^2}{N^2}\lp1-2  \,\frac{C^2}{A^2+B^2}\rp
\bar\beta\right. \nonumber\\
&&\qqq\qqq\qqq\qqq\qqq\qqq\qqq\ \ \left.+
\lp 1-2\,\frac{C^2}{N^2}\,\rp\eta _{-1}+
\lp-2+3  \,\frac{C^2}{N^2}\,\rp
\bar\eta \rb,
\ee
\bb
S _0&\approx &\epsilon\delta \,\frac{A^2+B^2}{N} \,\frac{a_0^2}{a_{-1}}\, \chi ^2
\nonumber\\
&&\qqq\ \ 
\cdot\lb 1
+
\lp\,\frac{A^2}{N^2}\,-3\,\frac{B^2}{N^2}\,
+\,\frac{A^2-B^2}{A^2+B^2}\,\frac{C^2}{N^2}\,\rp\beta_{-1}\,
-2\lp1-2\,\frac{B^2}{N^2}\,\rp\bar\beta
\right. \nonumber\\
&&\qqq\qqq\qqq\qqq\qqq\qqq\qq\left.
+\lp1-2\,\frac{C^2}{N^2}\,\rp\eta_{-1}\,
-2\lp1-2\,\frac{C^2}{N^2}\,\rp\bar\eta
\rb,\\[3mm]
\Omega  _{-1}&\approx &\frac{\epsilon\delta}{4\pi} \,\frac{A^2+B^2}{Na_{-1}} \,\lb1
+\lp \,\frac{B^2}{N^2}\,-2 \,\frac{B^2}{A^2+B^2}\, \rp\beta_{-1}
+\,\frac{C^2}{N^2}\, \eta_{-1}
\rb,\\[3mm]
\ell&\approx &\frac{L}{4\pi \chi ^2a_0^2} \,\frac{a_{-1}^2}{a_0^2}
\lb1
-\lp1-5\,\frac{B^2}{N^2}\,\rp\beta_{-1}\,
+2\lp1-2\,\frac{B^2}{N^2}\,\rp\bar\beta
\right. \nonumber\\[3mm]
&&\qqq\qqq\qqq\ \ \left.
-\lp1-5\,\frac{C^2}{N^2}\,\rp\eta_{-1}\,
+2\lp1-2\,\frac{C^2}{N^2}\,\rp\bar\eta\,
\rb.\label{apparent}
\ee
In linear approximation, the formula by Koivisto \& Mota (2008b) reads:\\
$
\ell\approx \frac{L}{4\pi \chi ^2a_0^2} \,\frac{a_{-1}^2}{a_0^2}
\lb1
+2\,\frac{B^2}{N^2}\,\beta_{-1}
+2\,\frac{B^2}{N^2}\,\bar\beta
+2\,\frac{C^2}{N^2}\,\eta_{-1}
+2\,\frac{C^2}{N^2}\,\bar\eta
\rb.$

\subsection{Dynamics}

For the dynamics we will not only suppose $\beta(t)$ and $\eta(t)$ small, but also $\beta'(t)/ \sqrt{3\Lambda }$ and $\eta'(t)/ \sqrt{3\Lambda }$ small and indicate by $\approx$  the leading approximation in all small quantities.

From equations (\ref{Hy}) and  (\ref{Hz}) we get:
\bb
L=V(H_y-H_x)=V\left(\frac{b'}{b}-\frac{a'}{a}\right)=V\frac{\beta'}{1+\beta}\approx V\beta'=V_0\,\beta'_0,\\[3mm]
M=V(H_z-H_x)=V\left(\frac{c'}{c}-\frac{a'}{a}\right)=V\frac{\eta'}{1+\eta}\approx V\eta'=V_0\,\eta'_0.
\ee
By its definition (\ref{defxi}), the bifurcation parameter becomes
\bb \xi =\,\frac{2}{g}\sqrt{\,\frac{\beta _0^{'2}-\beta '_0\eta'_0+\eta _0^{'2}}{3\Lambda }\, } \,\ll 1 ,\qq{\rm with}\qq
g\dpp=\,\frac{8\pi G\,\rho _0}{\Lambda} \,,
\ee 
showing that we are in the first case of the bifurcation. Linearising the scale factors in this case we get:
\bb \frac{a}{a_0}&\approx&\lp\,\frac{g}{2} \,\rp^{1/3}\lp \cosh u-1\rp^{1/3}
\lb1+{\textstyle\frac{4}{3}} \,\frac{\beta'_0+\eta'_0}{g\,\sqrt{3\Lambda }}\,  \la\,\frac{1}{e^u-1}\, -{\textstyle\frac{1}{2}} \lp
\sqrt{g+1}\,-1\rp\ra\rb,
\\[3mm]
\frac{b}{b_0}&\approx&\lp\,\frac{g}{2} \,\rp^{1/3}\lp \cosh u-1\rp^{1/3}
\lb1+{\textstyle\frac{4}{3}} \,\frac{-2\beta'_0+\eta'_0}{g\,\sqrt{3\Lambda }}\,  \la\,\frac{1}{e^u-1}\, -{\textstyle\frac{1}{2}} \lp
\sqrt{g+1}\,-1\rp\ra\rb,
\\[3mm]
\frac{c}{c_0}&\approx&\lp\,\frac{g}{2} \,\rp^{1/3}\lp \cosh u-1\rp^{1/3}
\lb1+{\textstyle\frac{4}{3}} \,\frac{\beta'_0-2\eta'_0}{g\,\sqrt{3\Lambda }}\,  \la\,\frac{1}{e^u-1}\, -{\textstyle\frac{1}{2}} \lp
\sqrt{g+1}\,-1\rp\ra\rb,
\ee
and for the eccentricities:
\bb
\beta&\approx&-4 \,\frac{\beta'_0}{g\,\sqrt{3\Lambda }}\,  \la\,\frac{1}{e^u-1}\, -{\textstyle\frac{1}{2}} \lp
\sqrt{g+1}\,-1\rp\ra ,\\[3mm]
\eta&\approx&-4 \,\frac{\eta'_0}{g\,\sqrt{3\Lambda }}\,  \la\,\frac{1}{e^u-1}\, -{\textstyle\frac{1}{2}} \lp
\sqrt{g+1}\,-1\rp\ra .\ee
The consistency of these relations can be checked by computing its derivative with respect to time $t$
and by using equation (\ref{fctV}) with $\xi=0$ and $t= t_0$: 
\beq
g \,(\cosh u_0-1)=2.
\eeq
Next we solve the redshift, equation (\ref{red}),
\bb 
z+1\,\approx \,\frac{a_0}{a(t)}\lb1-\,\frac{B^2}{N^2}\,  \beta(t)-\,\frac{C^2}{N^2}\,\eta(t)\rb
,\ee
for the departure time of the photon at the supernova. To alleviate notations, this departure time $t_{-1}$ is now simply written $t$ or will be omitted. In linear approximation we have:
\bb u=\sqrt{3\Lambda }\,(t-t_*)\approx u_F
+  \,\frac{2}{g\,\sqrt{3\Lambda }}
\lb\lp3\,\frac{B^2}{N^2}\,-1\rp\beta'_0
 +\lp3\,\frac{C^2}{N^2}\,-1\rp\eta'_0\rb
\,\frac{D}{\sqrt{g\,(z+1)^3+1}}\,,
\ee\bb
{\rm with}\qq u_F\dpp={\rm Arcosh}\lb \,\frac{2}{g\,(z+1)^3} \,+1\rb
\qq{\rm and}\qq D\dpp=\sqrt{g\,(z+1)^3+1}-\sqrt{g+1}
 \ee
For the apparent luminosity, 
\bb
\ell&\approx &\frac{L}{4\pi \chi ^2a_0^2} \,\frac{a_{-1}^2}{a_0^2}
\lb1
-\lp1-5\,\frac{B^2}{N^2}\,\rp\beta_{-1}\,
+2\lp1-2\,\frac{B^2}{N^2}\,\rp\bar\beta
\right. \nonumber\\[3mm]
&&\qqq\qqq\qqq\ \ \left.
-\lp1-5\,\frac{C^2}{N^2}\,\rp\eta_{-1}\,
+2\lp1-2\,\frac{C^2}{N^2}\,\rp\bar\eta\,
\rb,\ee
 we need the geodesic distance $\chi $, the scale factor $a$, and the eccentricities $\beta $ and $\eta$, all four at the time of departure of the photon from the supernova and we need the eccentricities averaged over the time of flight of the photon between the supernova and the Earth $\bar \beta$ and $\bar \eta$, all six as functions of redshift in linear approximation. Let us denote isotropic quantities, $\beta =\eta=0$, with a subscript $\cdot_F$ for Friedman: 
\bb \chi _F=\,\frac{I}{a_0}\, \sqrt{\frac{3}{\Lambda }},\ a_F=\,\frac{a_0}{z+1}\, ,\  \ell_F=\,\frac{L}{4\pi} \,\frac{\Lambda }{3I^2(z+1)^2}\, ,\ 
H_F( z)=\sqrt{\frac{\Lambda }{3}}\,\sqrt{g\,(z+1)^3+1},\ee
with the elliptic integral, cf appendix 1:
\bb
I(z)\dpp=\int_0^z\,\frac{\de \tilde z}{\sqrt{g\,(\tilde z+1)^3+1}}\, =\,\sqrt{\,\frac{\Lambda }{3}\,} \int_0^z\,\frac{\de \tilde z}{H_F(\tilde z)}.\ee
With these notations we can write:
\bb
\chi &\approx&\chi _F\lb
1+{\textstyle\frac{2}{3}} \,\frac{\beta'_0}{g\,\sqrt{3\Lambda }}
\la \sqrt{g+1}-\,\frac{z}{I}\, 
+
\lp1-3\,\frac{B^2}{N^2}\,\rp\,\frac{z+1}{I}\, \frac{D}{\sqrt{g\,(z+1)^3+1}}\ra
\right.\\&&\qqq\qq\ \ \left.
{\textstyle\frac{2}{3}} \,\frac{\eta'_0}{g\,\sqrt{3\Lambda }}
\la \sqrt{g+1}-\,\frac{z}{I}\, 
+
\lp 1-3\,\frac{C^2}{N^2}\,\rp\,\frac{z+1}{I}\, \frac{D}{\sqrt{g\,(z+1)^3+1}}\ra
\rb,\\[2mm]
a&\approx&a_F\lb
1+ \,\frac{2}{g\,\sqrt{3\Lambda }}\lp
\,\frac{B^2}{N^2}\,\beta'_0+\,\frac{C^2}{N^2}\,\eta'_0
\rp D\rb,\\[2mm]
\beta&\approx&-2\,\frac{\beta'_0}{g\,\sqrt{3\Lambda }}\,D, \label{bet}\qqq  
\bar\beta\approx-2\,\frac{\beta'_0}{g\,\sqrt{3\Lambda }}
\la \,\frac{z}{I}\, -\,\sqrt{g+1}\ra,
\\[2mm]
\eta&\approx&-2\,\frac{\eta'_0}{g\,\sqrt{3\Lambda }}\,D,\qqq
\bar\eta\approx-2\,\frac{\eta'_0}{g\,\sqrt{3\Lambda }}
\la \,\frac{z}{I}\, -\,\sqrt{g+1}\ra.
\ee
Finally, the  apparent luminosity as a function of redshift is in linear approximation:
\bb
\ell&\approx&\ell_F\lb 1-
 \,\frac{2}{\sqrt{3\Lambda }}
	\la
	\lp 1- 3\,\frac{B^2}{N^2}\, \rp\beta '_0\,
+\lp 1- 3\,\frac{C^2}{N^2}\, \rp\eta '_0\,
	\ra Q\,\rb,\\[2mm]
Q&\dpp=&
\,\frac{1}{3g}\,
 \la 4\lp \,\frac{z}{I}\,- \sqrt{g+1}\,\rp+\lp2\,\frac{z+1}{I\sqrt{g \,(z+1)^3+1}}\, -3\rp D\ra.
\ee
Note that for small redshift, we have 
\bb
 Q&\approx& \,\frac{1}{\sqrt{g+1}}\,
 ,\\[2mm]
\lim_{z\rightarrow 0}\lp z^2\ell\rp&\approx&\,\frac{L}{4\pi}\,H_{F0}^2\lb 1-{\textstyle\frac{2}{3}}  \lp 1- 3\,\frac{B^2}{N^2}\, \rp\,\frac{\beta'_0}{H_{F0}}\, 
-{\textstyle\frac{2}{3}}  \lp 1- 3\,\frac{C^2}{N^2}\, \rp\,\frac{\eta'_0}{H_{F0}}\, 
\rb.
\ee
Note also that Einstein's equations imply:
\bb H_x\approx H_F-{\textstyle\frac{1}{3}}\beta'-{\textstyle\frac{1}{3}}\eta'\,,\qqq
  H_y\approx H_F+{\textstyle\frac{2}{3}}\beta'-{\textstyle\frac{1}{3}}\eta'\,,\qqq H_z\approx
  H_F-{\textstyle\frac{1}{3}}\beta'+{\textstyle\frac{2}{3}}\eta'\,,\label{stretchdef} \ee
  and
  \bb \Omega _m+\Omega _\Lambda =1\,+O(\beta,\eta )^2,\qqq{\rm with}\qq
  \Omega _m\dpp = \,\frac{8\pi G\,\rho _0}{3H_{F0} ^2} \,,\qq
   \Omega _\Lambda \dpp = \,\frac{\Lambda }{3H _{F0} ^2} \,.
\ee  
 We have two privileged perpendicular directions, the $x$- and $z$-axes. The $y$-axis is then determined up to a sign, which is irrelevant because of reflection invariance of the Bianchi I metric (\ref{bianchi}).
We denote by $\theta   \in [0,\pi ]$ the angle of the incoming photon with the $z$-axis and by $\varphi  \in [0,2\pi )$ the angle between the projection of the incoming photon into the $xy$-plane and the $x$-axis:
\bb
A&=&N\cos\varphi  \,\sin\theta  ,\\
B&=&N\sin\varphi  \,\sin\theta  ,\\
C&=&N\qqq \,\cos\theta  .\ee
In these notations the  apparent luminosity as a function of redshift reads:
\bb
\ell&\approx&\ell_F\lb 1+
 \,\frac{2}{\sqrt{3\Lambda }}\la
	\beta'_0\lp 3\sin^2\varphi  \,\sin ^2\theta   -1\rp
	+\eta'_0\lp 3\cos ^2\theta   -1\rp\ra\,Q\,
	\rb,\label{liminosity} \ee
	with the isotropic apparent luminosity
	\bb
	\ell_F=\,\frac{L}{4\pi} \,\frac{\Lambda }{3I^2(z+1)^2}\, ,\label{lisotropic}
	\ee
the elliptic integral
\bb
I(z)\dpp=\int_0^z\,\frac{\de \tilde z}{\sqrt{g\,(\tilde z+1)^3+1}}\, ,\qqq g\dpp=\,\frac{8\pi G\,\rho _0}{\Lambda}, \ee
	and the auxiliary function
	\bb
	Q\dpp=
\frac{1}{3g}
 \la 4\lp \frac{z}{I}\,- \sqrt{g+1}\rp+\lp2\,\frac{z+1}{I\sqrt{g \,(z+1)^3+1}} -3\rp \!\!\lp\sqrt{g\,(z+1)^3+1}-\sqrt{g+1}
\rp\ra.\nonumber\\	\ee

\section{Data analysis}

To confront the Bianchi I metric with data we use the type 1a supernovae Hubble
diagram from the Union 2 sample  (Amanullah et al. 2010)
 with 557 supernovae up to a redshift of
1.4 and the Joint Light curve Analysis (JLA)  (Betoule et al. 2014) with 740 supernovae up to
a redshift of 1.3.  Note that 258  supernovae belong simultaneously to both samples. Supernovae celestial coordinates are obtained
from the SIMBAD astronomical database.

For the Union 2 sample, published data are the supernovae magnitudes at
maximum of luminosity corrected by time stretching of the light curve and
color at maximum brightness. The associated statistical and systematical
errors are provided by the full covariance matrix of supernovae magnitudes
including correlations.

The JLA published data provide the observed uncorrected peak magnitudes ($m_{\rm peak}$), time
stretching ($X1$) and color ($C$) with the full statistical and systematic covariance
matrix between all measurements including correlations between
supernovae. The total likelihood or $\chi^2$ is computed following the JLA
paper prescription and a rewriting of the likelihood computation program
provided by the COSMOMC package (Lewis \& Bridle 2002).
We choose to use the frequentist statistic  (Amsler et al. 2008)  based on $\chi^2$
minimization. The MINUIT package is used to find the minimum and to
compute errors with the second $\chi^2$ derivative. All presented results are
obtained after marginalization over nuisance parameters. The $\chi^2$
expression reads:
\bb \chi^2 = \Delta M^T V^{-1} \Delta M, \ee
where $\Delta M$ is the vector of differences between reconstructed ($m_r$) and expected
($m_e$) magnitudes at maximum and $V$ is the full covariance matrix including systematic
errors. The reconstructed magnitude for JLA reads:
\bb m_r = m_{\rm peak} + \alpha_s X1 - \beta_c C, \ee
where $\alpha_s$  and $\beta_c$ are fitted simultaneously with the additional free parameters. The expected
magnitude is written as $m_e(z) = m_s - 2.5 \log \ell(z)$ where $m_s$ is a normalization parameter
fitted to the data.

Let us define the three dimensionless  Hubble
stretch parameters today as \bb
h_{i} \dpp = H_{i0}/H_{F0}-1\qq {\rm for} \qq i={x,y,x}. \ee
Using formula (\ref{stretchdef}) they read:
\bb h_{x}  \approx -{\textstyle\frac{1}{3}}(\beta'_0+\eta'_0)/H_{F0}\,,\qq
    h_{y}  \approx  {\textstyle\frac{1}{3}}(2\beta'_0-\eta'_0)/H_{F0}\,,\qq 
    h_{z}  \approx {\textstyle\frac{1}{3}}(-\beta'_0+2\eta'_0)/H_{F0}\,, \label{hubblestretch} \ee
and verify $h_{x}+h_{y}+h_{z} = 0$.

The  apparent luminosity  $\ell(z)$ is computed with equation
(\ref{liminosity}) rewritten in terms of the two Hubble
stretch parameters $h_{z},\, h_{x}$.
\bb
\ell&\approx&\ell_F\lb 1+
 \,\frac{2}{ \sqrt{\Omega_{\Lambda} }}\la
	h_{z}\lp \cos^2 \theta -\sin^2\varphi \, \sin^2 \theta  \rp
	+h_{x}\, \cos 2\varphi \, \sin^2 \theta \ra\,Q\,
	\rb, \label{mylum} \ee
We search for preferred directions  by scanning the celestial sphere in steps
of 1 degree in right ascension and declination. For each direction ($r_z$,
$d_z$) assumed to be the $z$ direction, we minimize the $\chi^2$ over all other free parameters including
$\Omega_m$ and assuming a flat Universe. The minimum $\chi^2$ in the ($r_z$,$d_z$)
plane defines the privileged direction as well as the confidence level
contours. 

Because of the huge number of $\chi^2$ to minimize and because
the JLA sample requires many large matrix inversions, the following
analysis represents more than 100\,000 hours of computing time on a single CPU. To
speed up the processing, we use the EGEE (Enabling Grids for E-sciencE) datagrid facility with the
DIRAC web interface (Tsaregorodtsev et al. 2008).

\subsection{Bianchi I with axial symmetry}

We first look for a unique preferred direction along $\vec u_z$. For the case of
axial symmetry, we take the Hubble stretch parameters in the $x$- and $y$-directions to be equal: $h_{x} = h_{y} = -h_{z}/2$.
Then equation (\ref{mylum}) simplifies to:
\bb
\ell&\approx&\ell_F\lb 1+
 \,\frac{1}{ \sqrt{\Omega_{\Lambda} }}\la
	h_{z}\lp 3 \cos^2 \theta - 1  \rp \ra\,Q\,
	\rb, \label{lumaxial} \ee
For each 
celestial direction $\vec u_z$, we define $\theta$ to be  the angle between $\vec u_z$ and the direction of the incoming photons from the
supernovae. The isotropic apparent  luminosity is computed by rewriting
formula (\ref{lisotropic}) as:
	\bb
	\ell_F=\,\frac{L H_{F0}^2}{4\pi} \,\frac{1-\Omega_m }{I(z)^2(z+1)^2}\,
	\ee
The unknown constant term $L H_{F0}^2 / 4 \pi$ is absorbed into the fitted
normalization parameter $m_s$. The elliptic integral $I(z)$ is computed numerically.
 
\begin{figure}[h]
\begin{center}
\epsfig{figure=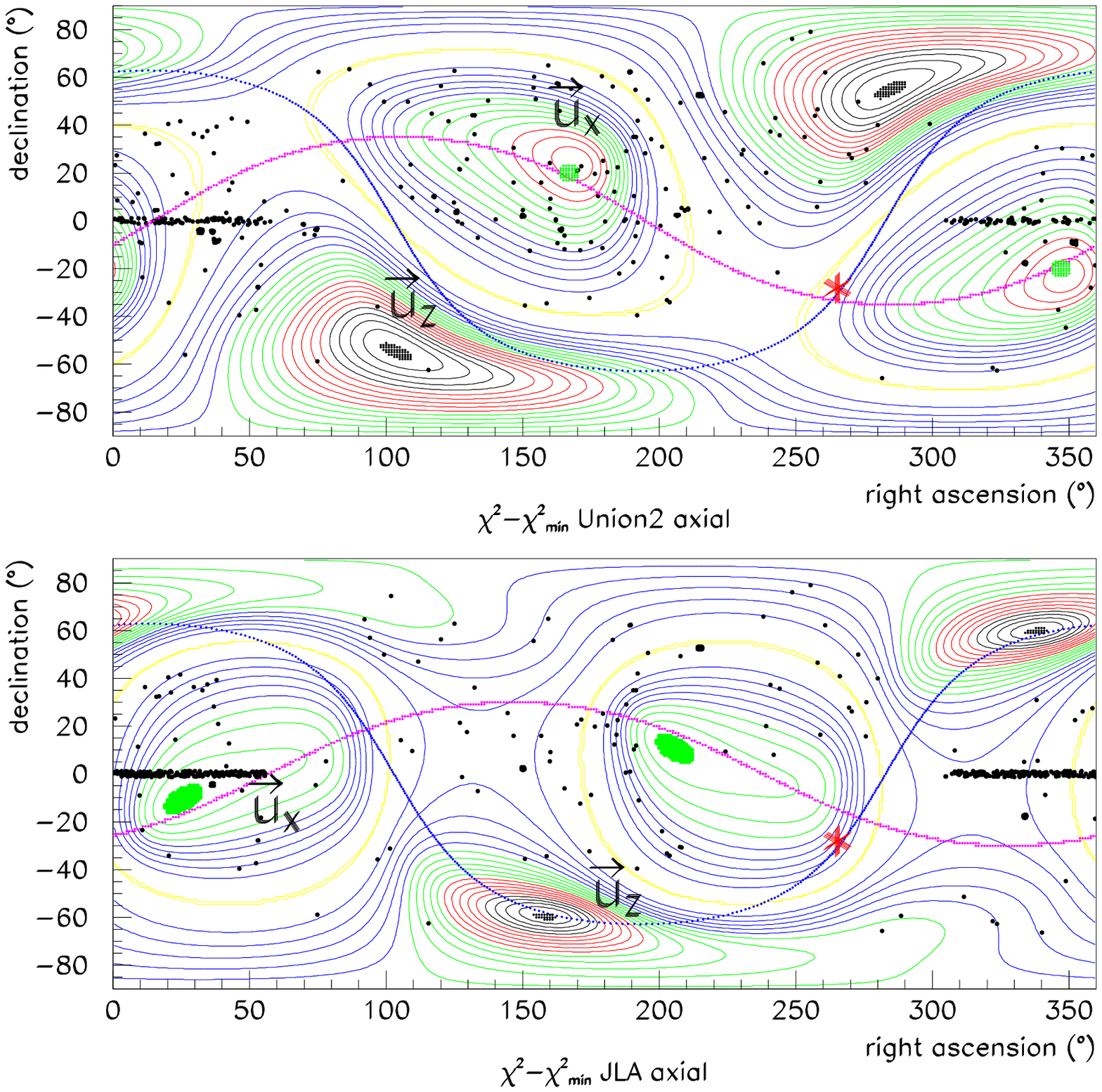,width=15cm}
\caption[]{Confidence level contours of privileged directions in arbitrary color
  codes for Bianchi I spacetimes with axial symmetry. Black points represent supernova positions. Note the accumulation  of supernovae in the equatorial plane. The blue line is the
  galactic plane and the purple line is the plane transverse  to the main privileged
  direction $\vec u_z$ (gray speck). The green specks show the secondary
  directions. The red star is the direction towards our  galactic center.}
\label{fig1}
\end{center}
\end{figure}

Figure \ref{fig1} shows the confidence level contours with  arbitrary color
codes around the eigendirections on the celestial sphere for the Union 2 and
JLA samples. The black points show the supernovae positions.
Notice that the Bianchi I metric is symmetric under space reflections. We therefore 
 expect two back-to-back eigendirections with the same eigenvalues, which is indeed the case. Figure \ref{fig1} has two distinct eigendirections, a main one indicated by a gray speck and a secondary one in green.
The full blue line is the galactic plane and the purple one
corresponds to the plane orthogonal to the main eigendirection (gray
speck). The statistical significance of the main eigendirection for the Union 2 sample is about
$41\%$ while it increases to $58\%$ for JLA.   The main eigendirections in the two samples are statistically
compatible (Table~\ref{table1}). They are close to the
galactic plane and almost orthogonal to the direction from the sun to the galactic center. For both
samples, the Hubble stretch $ h_z $ is negative and at most 1.2
$\sigma$ away from zero.

\begin{table}[htbp]
\begin{center}
\begin{tabular}{||c||c|c||c|c||} \hline
  Sample                 &   \multicolumn{2}{c||}  {Union 2}   & \multicolumn{2}{c||}{JLA}  \\  \hline
  Direction              &      main       & secondary       & main            & secondary  \\ \hline
right ascension($^\circ$)    &$103 \pm 58$     & $169 \pm36$     &  $158 \pm 29$   &   $160\pm97$    \\  \hline
declination ($^\circ$)       &$-56 \pm 36$     & $24 \pm 35$     &  $ -60 \pm 11$  & $-70 \pm50$     \\  \hline
galactic longitude ($^\circ$)&$88 \pm 45$      & $36 \pm 87$     &  $106 \pm 16$   &   $23 \pm 41 $\\  \hline
galactic latitude ($^\circ$) &$22\pm26$         & $-68 \pm 30$    &    $1 \pm 10$  & $19 \pm 31 $   \\  \hline
Hubble stretch in $\%$    &$-1.1\pm 1.1$    &$0.7 \pm 0.9$    &$-1.7 \pm 1.3$   & $0.5 \pm 0.9$\\  \hline
$\chi^2$ Bianchi I       &$529.64$         & $529.98$        &     $699.35$    & $700.43$             \\  \hline
$\Omega_m$               &\multicolumn{2}{c||}{$0.27\pm  0.04$}&\multicolumn{2}{c||}{ $0.29\pm 0.03$}\\  \hline \hline
$\chi^2$ isotropic       &\multicolumn{2}{c||}{ $530.71$}     & \multicolumn{2}{c||} {$701.3$}   \\ \hline
\end{tabular}
\caption[]{Fit results (1$\sigma$ errors) for Bianchi I with axial symmetry. We have used the invariance of the Bianchi I metric under space reflection to constrain the eigendirections to the hemispheres with right ascension or galactic longitude between $0^\circ$ and $180^\circ$.}
\label{table1}
\end{center}
\end{table}

 In Figure \ref{fig1} the green specks represent a
secondary direction with a weaker statistical significance ($30\%$ and $35\%$
confidence level for Union 2 and JLA respectively). They are in the plane
orthogonal to the main direction (purple line). The Hubble stretch in
this secondary direction has an opposite sign. Even though it is a weak
statistical effect, data seem to call for a second eigendirection.  This
feature is clearly observed in Figure \ref{fig3} presenting our simulation program. As a
possible consequence, the minimum $\chi^2$ is not as good as one would expect
from three additional degrees of freedom compared to the isotropic case. We conclude that 
there is some stress in the data when confronted to the Bianchi I with axial symmetry.

\subsection{Tri-axial Bianchi I}

The last remarks motivate us to extend the fit to the general  Bianchi I metric with three distinct eigenvalues and three orthogonal eigenvectors. 

As in the previous subsection we start with a first  eigendirection $\vec u_z$ defined
by its celestial coordinates ($r_z$,\,$d_z$):
 \bb \vec u_z \dpp= (\cos r_z \cos d_z,\, \sin r_z \cos d_z,\, \sin d_z) \ee
We then construct an orthonormal basis ( $\vec v_x,\,\vec v_y,\,\vec v_z = \vec u_z$) as follows: 
We choose the second unit vector $\vec v_x \dpp= (\sin
r_z,-\cos r_z,0)$ orthogonal to $\vec v_z$. The third unit vector is then defined by:
 $\vec v_y \dpp= \vec v_z \wedge \vec v_x$.

The second eigendirection $\vec u_x$ is
 obtained by rotating the vector $\vec v_x$ by an angle $\gamma$ around
 $\vec u_z$. The third eigendirection is completely fixed as
 well as the corresponding stretch parameters. The  angle $\theta$ is the angle between $\vec u_z$ and the direction of the incoming
 photons from the supernovae and   $\varphi$ is the angle between
  $\vec u_x$  and the projection of the incoming photon into the
 ($\vec v_x$,\,$\vec v_y$) plane.

For each direction defined by its celestial coordinates we compute the 
$\chi^2$ with formula (\ref{mylum}) and minimize it with respect to the following set of free
parameters: $m_s$, $\alpha_s$, $\beta_c$, $\Omega_m$, $h_z$, $h_x$
and $\gamma$.

\begin{figure}[htbp]
\begin{center}
\epsfig{figure=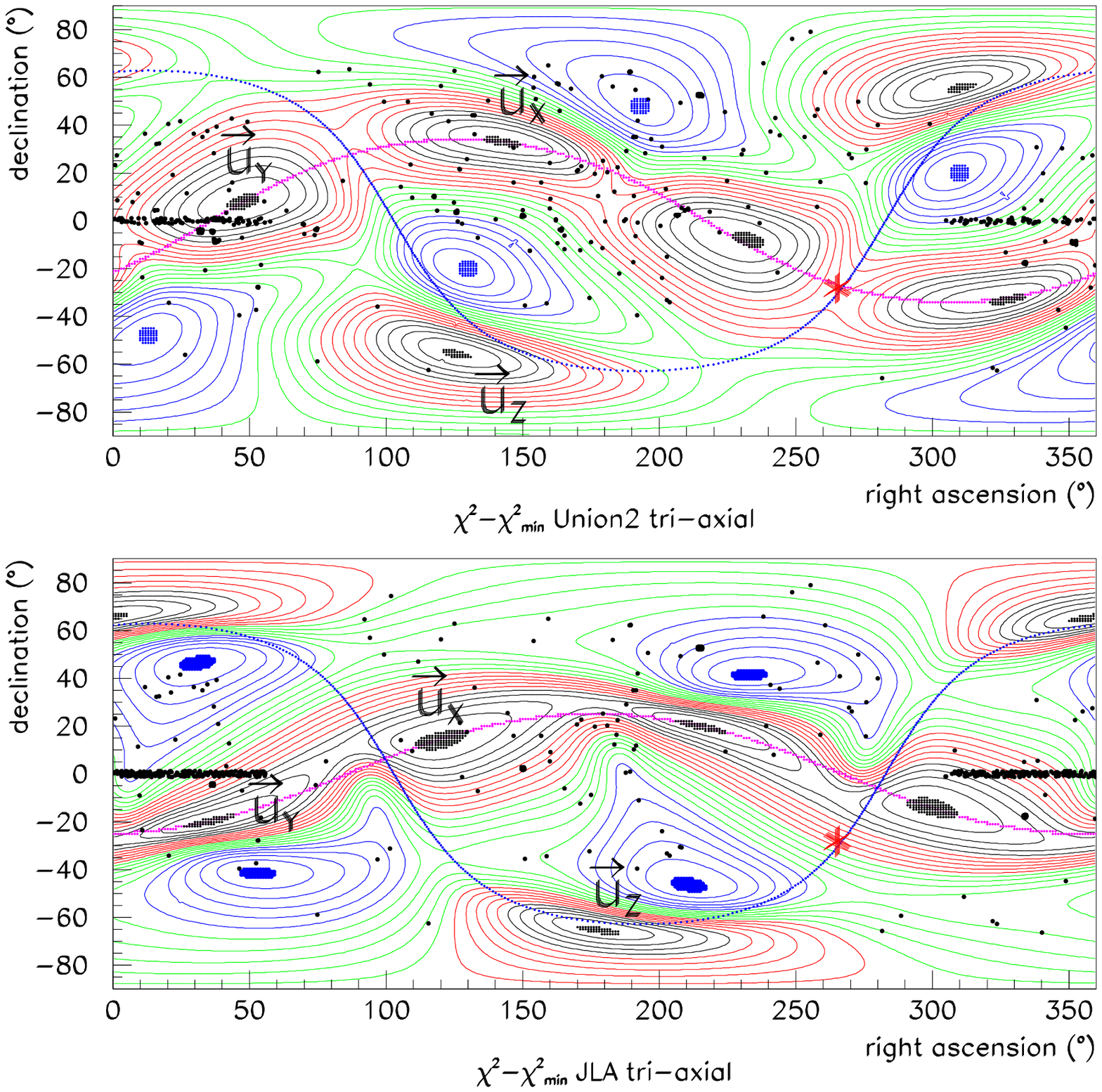,width=15cm}
\caption[]{Confidence level contours of privileged directions in  arbitrary color
 codes for the tri-axial Bianchi I metric. Black points represent supernova positions.  Note the accumulation of supernovae in the equatorial plane.
 The blue line is the
  galactic plane and the purple line is the plane transverse  to the main privileged
  direction $\vec u_z$ (gray speck). 
  Blue specks correspond to regions where
  $\chi^2$ is maximum. The red star is the direction towards our  galactic center.}
\label{fig2}
\end{center}
\end{figure}

Figure \ref{fig2} shows the confidence level contours with  arbitrary color
codes around the eigendirections on the celestial sphere. The  gray specks
mark the three eigendirections in each hemisphere. They materialize the unit vectors $\vec u_x$,
$\vec u_y$ and $\vec u_z$. The main eigendirection has by definition the largest Hubble stretch in absolute value and is given by $\vec u_z$. As in the axial case, this direction is 
close to the galactic plane (blue line).
The two remaining eigendirections are contained in the plane orthogonal to the main eigendirection $\vec u_z$
(purple curve).
 Blue specks within blue contours are regions where $\chi^2$ is {\it maximum}. They correspond to the
underprivileged regions that we expect on a sphere
with several minima of $\chi^2$.

Table \ref{table2} summarizes the eigendirections and eigenvalues for the Union 2
and JLA supernovae samples. The main eigendirections $\vec u_z$ of both samples are
close to each other. And they are statistically compatible with the main eigendirection of the axially symmetric case. The main Hubble stretch $h_z$ is negative
with a maximum significance of $1.3 \sigma$. In the JLA sample,
both secondary Hubble stretch parameters are very similar, which is compatible with the axial
symmetry. The improvement in the minimum $\chi^2$ compared to the isotropic case is of
the order of 2 units which is still too low for four added  degrees of freedom.

\begin{table}[htbp]
\begin{center}
\begin{tabular}{||c||c|c|c||c|c|c||} \hline
  Sample             &   \multicolumn{3}{c||}  {Union 2}   & \multicolumn{3}{c||}{JLA}  \\  \hline
  Direction          &   $\vec u_x$&$\vec u_y$    &   $\vec u_z$  &  $\vec u_x$ &  $\vec u_y$&   $\vec u_z$    \\ \hline
ascension($^\circ$)      &$ 142\pm 67$ &$ 48\pm59$    &  $ 126\pm 52$ &$160\pm30$  &$34 \pm 64$& $177 \pm56$   \\  \hline
declination($^\circ$)   &$33 \pm 21$  &$8 \pm 32$    &  $ -58 \pm 21$&$28\pm14$    &$-20 \pm 20$& $ -65\pm14$  \\  \hline
longitude($^\circ$)     &$12 \pm 35$  &$ 172\pm 51$  &  $91 \pm 30$  &$ 28\pm 28$    &$ 18\pm 36$& $ 116\pm25$ \\  \hline
latitude($^\circ$)      &$-47 \pm 52$ &$ -41\pm 53$  &  $10 \pm 20$  &$ -21\pm 60 $   &$ 69\pm 57$& $ 3\pm11$  \\  \hline
stretch($\%$)&$1.2\pm 1.1$ &$ 0.3 \pm 0.7$&$-1.5 \pm 1.3$ &$ 0.7\pm 1.0$&$1.1\pm 1.1$& $-1.8\pm1.4$ \\  \hline
$\chi^2$ Bianchi I   &\multicolumn{3}{c||}{$529.08$}     &\multicolumn{3}{c||}{$699.1$}     \\  \hline
$\Omega_m$           &\multicolumn{3}{c||}{$0.27\pm  0.04$}&\multicolumn{3}{c||}{ $0.28\pm 0.04$}\\  \hline \hline
$\chi^2$ isotropic   &\multicolumn{3}{c||}{ $530.71$}     & \multicolumn{3}{c||} {$701.3$}   \\ \hline
\end{tabular}
\caption[]{Fit results (1$\sigma$ errors) for tri-axial  Bianchi I. We have used the invariance of the Bianchi I metric under space reflection to constrain the eigendirections to the hemispheres with right ascension or galactic longitude between $0^\circ$ and $180^\circ$.}
\label{table2}
\end{center}
\end{table}

\subsection{Preliminary discussion}

The largest Hubble stretch we find in absolute value is equal to $h_z=-1.8\,\%$ with a
statistical significance of about $1.3 \sigma$. In the following sense, this corresponds to a pumpkin-like Universe in the future:

Consider a small, 2-dimensional, comoving sphere today, $t = t_0$, in a Bianchi I Universe with axial symmetry around the $z$-axis, $a(t)=b(t)$, and with negative Hubble stretch $h_z$. Recall that in our conventions $a(t_0) = b(t_0) = c(t_0)$. Then this sphere evolves with Einstein's equations in the future, $t > t_0$, into an oblate ellipsoid of revolution (pumpkin), $a(t) = b(t) > c(t)$. However, it comes from a prolate ellipsoid of revolution (rugby ball), $a(t) = b(t) < c(t)$ in the past, $t<t_0$. Indeed Einstein's equations imply that the Hubble stretches cannot change sign.
 
 The main privileged direction is in all cases contained
in the galactic plane and almost orthogonal to the direction between the sun and the galactic center. One is tempted to think that we are just measuring our proper velocity around the galactic center. In principle, the observer's proper velocity is already corrected for in the calibration of the data.
Nevertheless, we will test the hypothesis of a forward-backward asymmetry. To this end,
we split the supernovae of both samples into two hemispheres with respect to the main eigendirection $\vec u_z$,
backward ($\vec u_z^+$) and forward ($\vec u_z^-$) corresponding
respectively to the regions above and below the purple line in Figure
\ref{fig2}.  The number of supernovae per hemisphere is  338
  (respectively 219) in the backward (respectively forward) hemisphere for Union 2
 and 100 (respectively 640)  for JLA.

Using the axially symmetric fitting procedure and fixing 
$\vec u_z$ to the main direction as in Table \ref{table2}, we find 
$h_z^+= (1.3\pm2.1)\, \% \, ,h_z^-= -(2.3\pm1.5)\,  \%$ for Union 2 and\\ 
$h_z^+= (2.8\pm2.9)\, \% \, ,h_z^-= -(1.8 \pm1.5)\,  \%$ for JLA.

This small forward-backward asymmetry is no more than a $1.6 \sigma$
statistical effect and does not allow us to draw any relevant conclusion.

\section{Outlook}

\subsection{Comparison with other observations}

The Bianchi I metric is also used to decipher anisotropy in CMB data (Cea 2014) and in apparent proper motion measurements of extragalactic sources (Darling 2014). 

Cea (2014) uses the axially symmetric Bianchi I metric to fit the Planck and WMAP data and finds an eccentricity  at redshift $z=1090$ of $e=(0.86 \pm 0.14)\cdot 10^{-2}$ and an eigendirection with galactic latitude of $\pm 17^\circ$. In our notations we have,
\bb \beta =\sqrt{1-e^2}-1,\qq \eta = 0.\ee 
We use the first of equations  (\ref{bet}) to compute $\beta '_0$ and the second of equations (\ref{hubblestretch}) to get the main Hubble stretch.
Table \ref{table cea} compares Cea's results to ours.
Although Cea's Hubble stretch has the opposite sign and is smaller than ours by 8 orders of magnitude, the results are still compatible with each other.

\begin{table}[htbp]
\begin{center}
\begin{tabular}{|c|c|c|c|} \hline
  Sample                 &  CMB & {Union 2}   & {JLA}  \\  \hline

galactic longitude ($^\circ$)& &$88 \pm 45$        &  $106 \pm 16$      \\  \hline
galactic latitude ($^\circ$) &$\pm 17$&$22\pm26$          &    $1 \pm 10$     \\  \hline
Hubble stretch in \% & $(5.7\pm 1.8)\cdot 10^{-8}$    &$-1.1\pm 1.1$      &$-1.7 \pm 1.3$   \\  \hline

\end{tabular}
\caption[]{Cea's fit (Cea 2014) (1$\sigma$ errors) of Bianchi I with axial symmetry to CMB data compared with the two present fits.}
\label{table cea}
\end{center}
\end{table}

Darling (2014) uses a tri-axial Bianchi I metric to fit the apparent motion, `drift' for shortness, of 429 extragalactic radio sources measured by Titov \& Lambert (2013) using Very Long Baseline Interferometry. Table \ref{table darling} compares Darling's results to ours. His main Hubble stretch has the same sign as ours but is ten time larger. Although the results are again compatible statistically, their head-on comparison raises a conceptual difficulty. Unlike the Hubble diagram, drift measurements  do depend on the peculiar velocity (and acceleration) of the observer  and the separation of anisotropy induced by this peculiar velocity from anisotropic expansion at cosmological scale is delicate (Fontanini 2009). This is also true for measuring redshift distributions of quasars (Singal 2014).

\begin{table}[h]
\begin{center}
\begin{tabular}{||c||c|c|c||} \hline
  Sample             &\multicolumn{3}{c||}  {drift}    \\  \hline
 \hline
ascension($^\circ$)   &$ 102\pm 24$ &$ 13\pm15$    &  $ 11\pm 33$      \\  \hline
declination($^\circ$)   &$1\pm14$    &$-47 \pm 26$& $ 43\pm26$  \\  \hline
stretch($\%$)&$-19\pm 7$ &$ 17 \pm 7$&$2 \pm 7$  \\  \hline
 \hline
 
 Sample              &    \multicolumn{3}{c||}  {Union 2}    \\  \hline
 \hline
ascension($^\circ$)   &
  $ 126\pm 52$ & $ 322\pm 67$ &$ 48\pm59$       
\\  \hline
declination($^\circ$)   &
   $ -58 \pm 21$ & $-33 \pm 21$  &$8 \pm 32$    
\\  \hline
stretch($\%$) & $-1.5 \pm 1.3$  &
$1.2\pm 1.1$ &$ 0.3 \pm 0.7$
\\  \hline
 \hline
 
 Sample             &\multicolumn{3}{c||}{JLA}  \\  \hline
 \hline
ascension($^\circ$)   
& $177 \pm56$   & $34 \pm 64$   & $160\pm30$ 
\\  \hline
declination($^\circ$)   
& $ -65\pm14$  & $-20 \pm 20$  & $28\pm14$ 
\\  \hline
stretch($\%$) 
& $-1.8\pm1.4$ & $1.1\pm 1.1$  & $ 0.7\pm 1.0$
\\  \hline
 \hline
\end{tabular}
\caption[]{ Darling's fit (Darling 2014)  (1$\sigma$ errors) of tri-axial Bianchi I to the apparent proper motion (drift) of 429 extragalactic radio sources compared with the two present fits. For ease of comparison we have changed the sign of the second eigendirection in Union 2.}
\label{table darling}
\end{center}
\end{table}

\subsection{Future prospects}

It is fair to say that all three observations, CMB, drift and Hubble diagram,
pick up an intriguing signal of anisotropy when analysed in terms of a
Bianchi I cosmology. All three signals fail to be significant. All three
signals  have some tension with each other. However two signals can look forward to exciting new data in the near future.

After a successful launch in December 2013, the satellite Gaia has started its five-year period of data taking. These data should reduce the error bars on the Hubble stretch measured through the drift of quasars from the present seven percent to one percent (Darling 2014).

We count on the Large Synoptic Survey Telescope (LSST) (Abell et al. 2009) to reduce the error bars in the Hubble diagram.
LSST is a 6.7 meter telescope being built in Chile and that should start taking data in seven years. It will carry out a survey of $20\,000$ square degrees of
the sky, essentially the southern hemisphere, in six photometric bands with a main cadence for observation of 3 to 4
days allowing discovery and sampling of light curve supernovae up to a
redshift of about 0.8. The total number of SNe Ia in the main survey 
with a photometry sufficient for light curve fitting and photometric redshift
measurement is of the order of $50\,000$ per year.

Let us see to what extent LSST can improve the present analysis.
To this end, we simulate randomly $50\,000$ supernovae in $20\,000$ square
degrees with a redshift distribution centered around $z \approx 0.45$ and
going up to $z \approx 0.8$. The magnitude error including intrinsic dispersion and
photometric light curve fitting error is taken to be 0.12. The redshift error
is set to $\sigma_z = 0.01\,(1+z)$  and is propagated to magnitude error.

Table~\ref{table3} shows expected errors for 1 year and 10 years of LSST
modelled by the tri-axial Bianchi I metric. Fiducial values for the main Hubble stretch
parameter are taken to be $5 \sigma$ away from zero for 1 year of LSST survey,
and $3\sigma $ respectively $1 \sigma $ in the two other  principLE directions. The main result is that after 10
years of LSST survey, stretch parameters can be estimated with an accuracy of
about $3 \cdot10^{-4}$ and the main eigendirection with an accuracy of a few degrees for a stretch parameter of the order
of $3 \cdot10^{-3}$.

\begin{figure}[htbp]
\begin{center}
\epsfig{figure=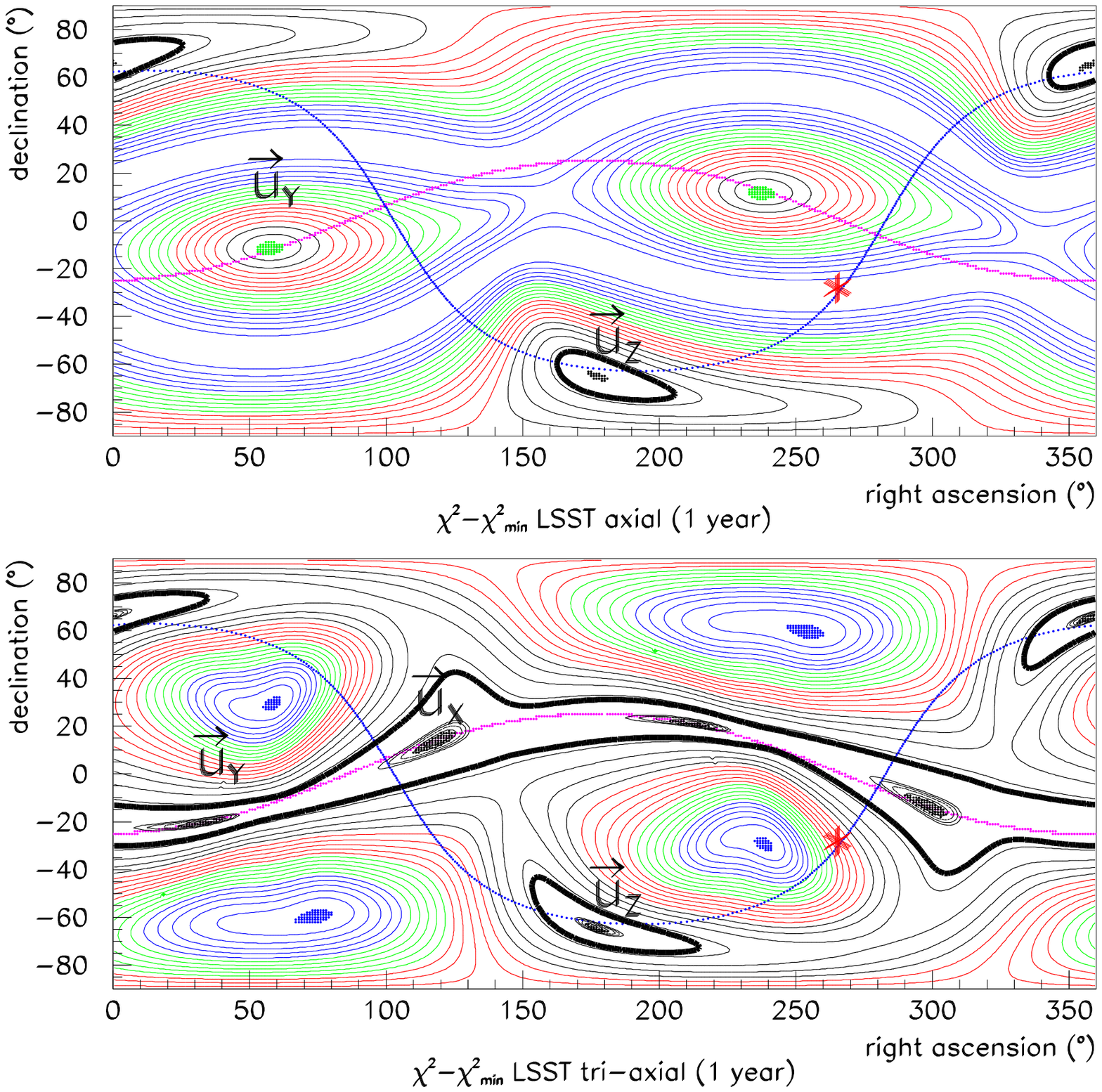,width=15cm}
\caption[]{Confidence level contours of privileged directions in  arbitrary color
 codes for the two fits using the axial and tri-axial Bianchi I metric  for one
 year of LSST. $68 \%$
   confidence level contours are drawn in bold. The blue line is the
  galactic plane and the purple line is the plane transverse to the main privileged
  direction $\vec u_z$ (gray speck). 
The red star is the direction towards our  galactic center. The green specks in the upper plot show the secondary
    directions in the axial fit while the blue specks in the lower plot show regions
    where $\chi^2$ is maximum.}
\label{fig3}
\end{center}
\end{figure}

Figure \ref{fig3} shows the confidence level contours for axial and tri-axial
Bianchi I metric fits with arbitrary color
codes around the eigendirections on the celestial sphere for 1 year of LSST.
 All features observed in these figures are similar to those observed on
  real data. Bold black contours mark $68\%$ confidence level
  and show the same kind of degeneracy as in figures \ref{fig2}.
 This is due to the very close values of the Hubble stretch parameters in
  the secondary eigendirections ($\vec u_x$,  $\vec u_y$).

\begin{table}[htbp]
\begin{center}
\begin{tabular}{||c||c|c|c|c||c|c|c|c||} \hline
  Sample             &   \multicolumn{4}{c||}  {LSST 1 year}   &  \multicolumn{4}{c||}{LSST 10 years}  \\  \hline
 Fitting method      &  \multicolumn{1}{c|} {Axial} & \multicolumn{3}{c||} {Tri-axial} &  \multicolumn{1}{c|} {Axial} & \multicolumn{3}{c||}{Tri-axial} \\ \hline
  Direction          &   $\vec u_z$ &   $\vec u_x$&$\vec u_y$    &   $\vec u_z$  &   $\vec u_z$  &$\vec u_x$ &  $\vec u_y$&   $\vec u_z$    \\ \hline
ascension($^\circ$)      &$ \pm 12$ &$ \pm 67$ &$ \pm 37$    &  $ \pm 20$ &$\pm4$   &$\pm20$  &$ \pm 12$& $ \pm 6$   \\  \hline
declination($^\circ$)    &$ \pm 7$  &$ \pm 21$  &$8 \pm 8$    &  $ \pm9 $ &$\pm2$    &$\pm7$    &$ \pm 3$& $ \pm3$  \\  \hline
stretch($\%$)        &$ \pm 0.06 $ &$ \pm 0.1 $ &$  \pm 0.1$&$ \pm 0.09$ &$ \pm 0.02$ &$ \pm 0.03$&$\pm 0.03$& $ \pm 0.03$ \\  \hline
\end{tabular}
\caption[]{1$\sigma$ errors for the two fits using  axial and tri-axial Bianchi I metrics 
  for 1 and 10 years LSST simulations with a tri-axial supernova distribution.}
\label{table3}
\end{center}
\end{table}

\section{Conclusions}

Today the Hubble diagram remains one of the cleanest windows on the Universe and still holds a lot of potential, both for data and theory.  Indeed the underlying theory is simply general relativity together with a possibly weakened cosmological principle and allows for precise calculations.  In view of the expected LSST data the Hubble diagram seems particularly well suited to separate geography from geometry in the sense of the Introduction.
\\
\vskip.3cm\noindent
{\bf \Large References}\\
\vskip.12cm\noindent
Abell P. A. et al., 2009,
     arXiv:0912.0201  [astro-ph.IM]
\\
Amanullah R., 
Lidman C., Rubin D., Aldering G., Astier P., Barbary K., Burns M.~S., 

Conley~A.
et al., 2010,  ApJ, 716, 712
\\
Amsler C. et al., 2008, Phys. Lett. B, 667
\\
Antoniou I., Perivolaropoulos L., 2010,
  JCAP, 1012, 12
  \\
  Betoule M. et al., 2014,  arXiv:1401.4064 [astro-ph.CO]
\\
  Campanelli L., Cea P., Fogli G. L., Marrone A., 2011,
  Phys.\ Rev.\ D,  83, 103503
  \\
   Campanelli L., Cea P., Fogli G. L., Tedesco L., 2011,
  Mod.\ Phys.\ Lett.\ A,  26, 1169
 \\
   Cea P., 2014,
   arXiv:1401.5627 [astro-ph.CO]
\\
Chang Z., Li X., Lin H.-N., Wang S., 2014a,
  Eur.\ Phys.\ J.\ C, 74, 2821
  \\
Chang Z., Li X., Lin H.-N., Wang S., 2014b,
  Mod.\ Phys.\ Lett.\ A, 29, 1450067
 \\
  Darling J., 2014,
  arXiv:1404.3735 [astro-ph.CO]
\\
Fontanini M., West E. J., Trodden M., 2009,
  Phys.\ Rev.\ D,  80, 123515
  \\
 Jimenez J. B., Salzano V., Lazkoz R., 2014,
   arXiv:1402.1760 [astro-ph.CO]
  \\
  Kalus B., Schwarz D. J., Seikel M., Wiegand A., 2013, A\&A, 553, A56
\\
 Koivisto T., Mota D. F., 2008a,
  ApJ, 679, 1
\\
Koivisto T., Mota D. F., 2008b,
  JCAP, 0806,018
\\
Kolatt T. S., Lahav O., 2001, MNRAS, 323, 859
\\
 Lewis A., Bridle S.,  2002,    Phys. Rev. D,  66, 103511, http://cosmologist.info/cosmomc/
 \\
M\'esz\'aros A., \v{R}\'ipa J., 2013,
  arXiv:1306.4736 [astro-ph.CO]
 \\
Quercellini C., Quartin M., Amendola L., 2009,
   Phys.\ Rev.\ Lett., 102, 151302
   \\
Saunders P. T., 1969, MNRAS, 142, 213
\\
Schwarz D. J., Weinhorst B., 2007, A\&A, 474, 717
   \\
Singal A, K., 2014,
  arXiv:1405.4796 [astro-ph.CO]
\\
Titov O., Lambert S., 2013, A\&A,
 559, A95
 \\
Tsaregorodtsev A. et al., 2008,    J. Phys.: Conf. Ser., 119, 062048, http://diracgrid.org
\\
  Yang X., Wang F. Y.,Chu Z., 2013,
    arXiv:1310.5211 [astro-ph.CO]
 \\
 Enabling Grids for E-sciencE,
    http://www.egi.eu
    \\
    The MINUIT-ROOT analysis package,
    http://root.cern.ch/drupal/
    \\
SIMBAD astronomical database:
    http://simbad.u-strasbg.fr/simbad/
  \\
\vskip.3cm\noindent
{\bf \Large Appendix: Elliptic integral}\\
\vskip.12cm\noindent
Consider the integral
\beq
I(z)=\int_0^z\,\frac{\de \tilde z}{\sqrt{g(1+\tilde z)^3+1}}\qq\qq z \geq 0.
\eeq
Its computation follows from M\'esz\'aros \& \v{R}\'ipa (2013).
The change of variables,
\bb
v=\frac 1{1+\tilde z},\ee
gives
\bb
 I(z)=\int_{(1+z)^{-1}}^1\frac{\de v}{\sqrt{v}\sqrt{g+v^3}},
\ee
followed by
\bb
v=g^{1/3}\,y,\qqq\qqq
 I(z)=g^{-1/3}\,\int_{g^{-1/3}(1+z)^{-1}}^{g^{-1/3}}\frac{\de y}{\sqrt{y}\sqrt{1+y^3}}.
\ee
Following M\'esz\'aros \& \v{R}\'ipa (2013) let us define
\beq
\cos\Big(\tht(z)\Big)\dpp=\frac{1+z-(\sqrt{3}-1)g^{-1/3}}{1+z+(\sqrt{3}+1)g^{-1/3}},
\eeq
which finally yields
\beq
I(z)=3^{-1/4}\,g^{-1/3}\Big(F(\tht(0),k)-F(\tht(z),k)\Big),\qq\qq k^2\dpp=\frac{2+\sqrt{3}}{4}<1.
\eeq
The elliptic integral (of first kind) is defined by
\beq
F(\phi,k)\dpp=\int_0^{\phi}\,\frac{\de u}{\sqrt{1-k^2\sin^2 u}}.\qq\qq 0<k^2<1.
\eeq

\end{document}